\documentclass{article}
\usepackage{iclr2027_conference, times}

\usepackage{times}
\usepackage[T1]{fontenc}
\usepackage[utf8]{inputenc}
\usepackage{amsmath,amssymb,amsthm}
\usepackage{graphicx}
\usepackage{booktabs}
\usepackage{microtype}
\usepackage{xcolor}
\usepackage[colorlinks=true,linkcolor=black,citecolor=black,urlcolor=black]{hyperref}
\usepackage{url}
\usepackage{placeins}
\usepackage{bm}

\newcommand{\R}{\mathbb{R}}
\newcommand{\E}{\mathcal{E}}                 
\newcommand{\Cb}{\mathcal{C}}                
\newcommand{\Xset}{\mathcal{X}}              
\newcommand{\enc}{\mathrm{Enc}}
\newcommand{\dec}{\mathrm{Dec}}
\newcommand{\swapop}{\Sigma}

\newcommand{\class}[1]{\textsf{#1}}
\newcommand{\vh}{\bm{h}}                    
\newcommand{\vx}{\bm{x}}                    
\newcommand{\vw}{\bm{w}}                    
\newcommand{\hvw}{\hat{\bm{w}}}             

\title{Steering intonational tunes in text-to-speech\\ synthesis}

\author{Juan Pablo Vigneaux \& Jennifer Cole \\
Northwestern University\\
\texttt{\{jpvigneaux,jennifer.cole1\}@northwestern.edu} 
}

\iclrfinalcopy

\begin{document}
\maketitle
{\let\thefootnote\relax\footnotetext{Code: \url{https://github.com/jpvigneaux/steering-intonation-tts}}}

\begin{abstract}
Text-to-speech systems built on residual vector quantization often generate audio in two stages:
a transformer backbone emits the first level of each frame's token stack, and a smaller depth
model generates the rest autoregressively. We ask which components of such a model are responsible
for prosodic intonation, more precisely for generating the English \emph{nuclear tune} (for
instance, the falling pitch melody that distinguishes an assertion from a question's rising
melody), and whether interventions on those components can be used to control intonation.
Prompt-based control mostly fails; only a final question mark is a reliable lever. Interchange interventions in
CSM and S2 locate the tune: the backbone's hidden state carries it to the waveform through the
depth model, and the first level of the stack carries little or none of it. Steering the backbone
state with a difference-of-means vector computed on new material generated with a final period or
question mark moves the tune, but the push toward an assertion degrades the voice. Steering vectors
computed instead from the backbone states associated with human imitations of
theoretically-specified tunes, whose text carries no punctuation, and applied inside the generation
loop, move most recordings to the other condition in both systems (even when fitted without the
speaker being steered), at a small cost in predicted naturalness. These imitation-derived vectors
are aligned with the one obtained from the two punctuated conditions far beyond a shuffled-label
null, so the backbone state of a punctuated prompt already represents intonation. The steered
recordings broadly become falling or rising, according to the intent, but the edge tone moves more
coherently than the pitch accent, and the target tune cannot be selected precisely beyond this
broad tendency.
\end{abstract}

\section{Introduction}
\label{sec:intro}

Neural audio codecs based on residual vector quantization (RVQ) and variants thereof (e.g., Mimi's
split RVQ) represent an audio signal as a low-frequency stream of discretized representations; in
fact, with a whole hierarchy of these, meant to capture the signal at increasing levels of fidelity
\citep{zeghidour2021soundstream,defossez2023encodec}. Many generative speech models take these
codec-derived discrete representations as tokens and split the stack. We call a \emph{staged
hierarchical generator} a model in which an autoregressive backbone runs along time and emits the
first level in the hierarchy (level $0$), and a smaller network, conditioned on the backbone's
hidden state and on the level-$0$ token, fills the other levels of each frame along depth. CSM
\citep{sesame2025csm} and Fish Audio S2 \citep{fishaudio2026s2} are text-to-speech systems of this
kind, and Moshi \citep{defossez2024moshi}, an adaptation of the RQ-Transformer
\citep{lee2022rqtransformer}, is a speech-to-speech one; VALL-E \citep{wang2023valle} and HALL-E
\citep{halle2025} split the stack at the same level but fill the other levels non-autoregressively
from the level-$0$ tokens. The split is an efficiency measure, and its consequence
for control has not been measured. This paper asks about control of intonation: it is not clear if intonation is computed by the backbone
transformer or by the depth transformer (or both), if level $0$ already contains any meaningful
representation of intonation, and where exactly one should intervene to change or improve
intonational behavior. Using probing, \citet{vigneaux2026probing} show that quantization affects
the encoding of intonation, and that levels $1$ and $2$ encode it more faithfully than level $0$;
we ask whether a similar phenomenon shows up at the level of generation.

Our work focuses on the expression of intonational \emph{nuclear tunes} in
American English, which are the pitch melodies carried from the stressed syllable of the last
accented word of an \emph{intonational phrase} to the end of that phrase. Nuclear tunes
distinguish  a tune with falling pitch marking assertions, for instance,\textit{``It's Marilyn.''} from a tune with rising pitch  marking questions, \textit{``It's Marilyn?''}
\citep{pierrehumbert1980}. Tunes are phonological categories (abstract sound features) rather than
speaker properties, and they are conventionally written as sequences of high and low tonal
targets: a pitch accent on the stressed syllable, then a phrase accent and a boundary tone at the
edge. In this work, we implicitly elicit rising and falling tunes from the models, by prompting
them with a full stop (condition \class{A}, assertion) or a question mark (condition \class{I},
interrogation); this particular tune contrast is practically the only one that can be consistently
elicited by a single orthographic sign: in general, tunes depend on the whole communicative
situation in a complex way. Moreover, we use human imitations of the eight model tunes provided by
the autosegmental-metrical theory of English intonation \citep{cole2023} as a ground truth for tune
distinctions; although the theory prescribes eight tunes (technically, monotonal nuclear tunes),
the human imitations of these theoretically specified tunes only support a five-way distinction
(\S\ref{app:shapes}).

In preliminary experiments, we tried to elicit a varied inventory of tunes from a text-to-speech
system in three different ways: by using various kinds of style tags, by providing repeated
recorded examples of one tune, or by prompting the model with a lead-in that, in humans, would
constrain the intonation of its continuation (\S\ref{app:prompting}). Eighteen inline style tags
naming typical uses of a rising tune (\texttt{[questioning]}, \texttt{[very uncertain]}, \dots)
leave S2's pitch indistinguishable from the untagged baseline across ten voices, and none moves
the contour toward a rise (Figure~\ref{fig:style-tags}). Lead-ins do not move the contour either,
and a recorded demonstration of the tune, given as the voice prompt, moves it only partway.
Changing the final punctuation mark is the only prompt-side lever that works reliably: controlling
intonation beyond that single orthographic switch requires intervening on the computation.

Broadly, we are interested in speech models' capacity to produce the whole spectrum of tunes as
humans use them. Here we take a narrower perspective: since prompting does not yield the expected
variety of tunes, we use the \class{A}/\class{I} contrast, which does change the tune consistently,
to study the mechanisms that compute intonation inside staged hierarchical generators.

\paragraph{Contributions.}
\begin{enumerate}\itemsep2pt
\item We locate the intonational tune in two staged hierarchical generators by interchange
interventions. Swapping the audio tokens generated by the depth model between utterances produced
under the assertion or interrogation condition (with matched sentence and voice) transfers the
intonation, while swapping level $0$ transfers neither tune nor words (in S2, a small part of the
tune). Re-running the depth model with the other utterance's backbone state $\vh_t$ transfers the
tune, and with its level-$0$ tokens does not: $\vh_t$ carries the tune contrast to the waveform
through the depth model (\S\ref{sec:where}).
\item We show that a difference-of-means edit on $\vh_t$, with the steering vector computed from
backbone states of \emph{generated} material under both punctuation conditions over the final
word, steers
the tune offline (i.e., by only re-running the depth model at each frame after adding the steering
vector), and fails in one direction: a question driven toward an assertion suffers degraded voicing and
harmonicity (\S\ref{sec:offline-limits}).
\item We fit difference-of-means steering vectors on the backbone states computed while the model
\emph{reads} human imitations of theoretically-specified nuclear tunes, whose transcripts carry no
punctuation, and apply them inside the generation loop. They shift the pitch contour near the end
of the utterance (the edge tone) to the other condition's contours in most recordings of both
systems ($56$ to $72\%$), also when fitted
without the speaker being steered, while predicted naturalness remains very similar to that of the target
condition.  Despite the effect of steering on the final region of the pitch contour, the resulting contours often fail to reach the intended tune and differ from the tunes  elicited from punctuation (\S\ref{sec:human}).
\end{enumerate}

\section{Related work}
\label{sec:related}\label{app:related}

\paragraph{Causal localization.} In causal analysis, an interchange intervention runs a model on
one input with some internal variables set to the values they take on another input, and reads what
changes in the output \citep{geiger2021causal,geiger2022iit}; in the interpretability literature
this is a form of activation patching \citep{meng2022rome,heimersheim2024patching}, and an instance
of causal mediation analysis \citep{vig2020causal}; our mediators are the levels of the token stack
and the backbone state read by the depth model. SpeechTokenizer \citep{zhang2024speechtokenizer} combines the first level
of one utterance with the other levels of another to transfer a speaker's voice (their \S5.2).

\paragraph{Activation steering.} Adding a fixed vector to hidden states at inference changes what
a language model generates \citep{subramani2022steering,turner2023actadd,zou2023repe}; the vector is
usually a difference of means between contrasting inputs \citep{panickssery2024caa,li2023iti}, which
steers better than a probe's direction \citep{li2023iti,marks2024geometry}, and its success is read as
evidence that the attribute is
linearly represented \citep{park2024linear}. Prompting often matches or beats steering where both
apply \citep{wu2025axbench}, as we also find for the single tune contrast that we can prompt.

\paragraph{Steering speech and music generators.} Difference-of-means and learned steering
vectors control emotion in flow-matching and language-model text-to-speech systems
\citep{xie2025emosteer,zhou2026emoshift}, neutralize accent \citep{yang2026accent}, and change
tempo and timbre in music generation \citep{facchiano2025music}; sparse-autoencoder features of a
text-to-speech language model steer laughter, gender and speech rate \citep{koriagin2026sae}. Closest
to our question, \citet{wang2026cocoemo} and \citet{wang2026geometric} ask which
module of a hybrid text-to-speech system (a speech language model followed by a flow-matching
decoder) to steer for mixed emotions. These studies steer utterance-level attributes; we steer a
linguistic attribute, a cue for meaning located on one word.

\paragraph{Controlling prosody.} In SSML \citep{w3c2010ssml}, the common language of
text-to-speech control, intonation can only be manipulated through explicit acoustic parameters
such as pitch targets, ranges, contours and rates. This requires a tune to be defined as a
time-varying pitch trajectory, which presupposes knowing the right trajectory for a particular
speaker and sentence content, and ignores the evidence that speakers select from a categorical
inventory rather than a continuum of time-varying trajectories \citep{cole2023}. Natural-language prompts and tags describe style at the level of the utterance
\citep{guo2023prompttts,yang2024instructtts,lyth2024parler}, and QI-TTS transfers a questioning
intonation from reference speech \citep{tang2023qitts}. Probing studies find prosodic information
in self-supervised speech representations \citep{lin2022prosody}, in discrete tokens derived from
them \citep{onda2025prosody} and across spoken-language-model pipelines \citep{naber2026probing},
and intonational categories in the levels of RVQ codecs \citep{vigneaux2026probing}.

\section{Setup}
\label{sec:setup}

\paragraph{Staged hierarchical generators.} Write $q_t = (q_t^0, \dots, q_t^{K-1})$ for the stack
of codewords of frame $t$ and $\vh_t$ for the backbone's final-layer hidden state at that frame.
Codewords are sampled stochastically, frame by frame, from a distribution $p_{\theta,\phi,\psi}$
that a staged hierarchical generator factorizes as
\begin{equation}
p_{\theta,\phi,\psi}(q_t \mid q_{<t}, c) \;=\; \underbrace{p_\phi(q^0_t \mid \vh_t)}_{\text{L0-head}}
\;\cdot\; \underbrace{\textstyle\prod_{k\ge 1} p_\psi(q^k_t \mid q^{<k}_t, \vh_t)}_{\text{depth
model}}, \qquad \vh_t = f_\theta(q_{<t}, c),
\label{eq:staged}
\end{equation}
with $c$ the text and voice conditioning. Here $f_\theta$, the backbone, and $p_\psi$, the depth
model, are transformers with parameters $\theta$ and $\psi$, and $p_\phi$ is a shallow network, the
\emph{L0-head}: in both systems a single matrix multiplication followed by a softmax (in S2 the
matrix is the backbone's token-embedding matrix, whose weights are tied to the output). We study CSM
\citep{sesame2025csm} and Fish Audio S2 \citep{fishaudio2026s2}. CSM's codec is Mimi
\citep{defossez2024moshi}, which CSM runs with $K = 32$ levels at $12.5$~Hz; S2's is built on the Descript Audio
Codec \citep{kumar2023dac}, with $K = 10$ levels at $21.5$~Hz ($44.1$~kHz audio, $2048$ samples per
frame). Level $0$ is referred to in the
literature as a `semantic' level, which is aligned during training with the representations
generated by a neural audio encoder: Mimi distils WavLM \citep{chen2022wavlm} into it and S2's
codec distils w2v-BERT~2.0. Both systems follow SpeechTokenizer \citep{zhang2024speechtokenizer},
which distils HuBERT \citep{hsu2021hubert}. The two codecs differ in how the other levels relate to
level $0$: S2's level-$1$ tokens quantize the residual that level $0$ leaves (and each further level
the residual of the one before), while Mimi forms a separate RVQ stack over the whole frame, in
parallel with level $0$ (a \emph{split} quantizer; \S\ref{app:rvq}). CSM and S2 apply a
root-mean-square normalization
\citep{zhang2019rmsnorm} to the backbone's final hidden state before feeding it to the L0-head and
the depth transformer. Our steering vectors are applied before this normalization.

\paragraph{Material.}
\label{sec:material}
The human recordings are the Basic8 corpus of \citet{cole2023}, in which speakers hear a
resynthesized recording of each of eight theoretically-specified combinations of high and low
pitch accent, phrase accent and boundary tone (the \emph{model tunes}) and imitate it on short
sentences. A tune is written by its three tones in that order: \class{hll} (high accent, low
phrase accent, low boundary tone) is the paradigmatic fall of an assertion and \class{lhh} the
paradigmatic rise of a question (\S\ref{app:notation}). The recordings are labelled by the model
tune, and clustering the speakers' renditions yields five groups that correspond almost exactly to
a partition of the eight labels into five classes ($92\%$ of speaker-and-tune means fall in the
cluster of their label's class, \S\ref{app:shapes}). As prompts for audio
generation, we use the three carrier sentences the corpus uses for every speaker (\textit{He
modeled Harmony}, \textit{She remained with Madelyn}, \textit{They honored Melanie}) with
recordings from ten speakers (five female, four male, one non-binary), spanning the corpus's range
of median $F_0$. Each generated utterance is prompted with the voice of one of those speakers:
eight of that speaker's recordings, one per tune, in random order and excluding the sentence to be
synthesized, so that no tune is privileged. Each sentence is generated three times per voice (three
\emph{iterations}), each with its own draw of the eight prompt recordings and its own seed. Ten
speakers $\times$ three sentences $\times$ three iterations $\times$ two punctuation conditions give
$180$ utterances per system, the \emph{ten-speaker set}; an assertion and a question with the same
voice, sentence and iteration form a \emph{matched pair} ($90$ per system). All audio is
force-aligned \citep{mcauliffe2017mfa}, and every measurement is taken over the final word. The final word carries the nuclear tune in these sentences: in the generated
utterances, the pitch contours associated with the two punctuation conditions differ over the final
word and not before it, as in human imitations of the corresponding tunes (\S\ref{app:where-differ}).

\paragraph{What we measure.}
\label{sec:measures}\label{sec:criterion}\label{sec:alignment}\label{sec:families}
$F_0$ (the acoustic correlate of auditorily perceived pitch) is tracked with Praat's filtered autocorrelation \citep{boersma1993hnr} and converted to ERB
(equivalent rectangular bandwidth), a psychoacoustic scale on which equal differences are heard as approximately equal pitch
movements whatever a speaker's range \citep{hermes1991erb}. Over the final word we take the \emph{excursion} (mean $F_0$ level of the
last fifth of voiced points minus that of the first fifth), the \emph{offset} (the mean $F_0$
level of the last fifth alone), the \emph{voiced fraction} (the proportion of time steps to which
the pitch tracker can assign an $F_0$ value), and the harmonics-to-noise ratio (HNR;
\citealp{boersma1993hnr}), which falls as a voice becomes hoarse or breathy \citep{yumoto1982hnr}. Over the whole utterance we measure predicted
naturalness with UTMOS \citep{saeki2022utmos} and speaker similarity to the voice prompt with
WavLM speaker embeddings \citep{chen2022wavlm} (\S\ref{app:quality}). A recording \emph{falls} or
\emph{rises} according to the sign and magnitude of its excursion, with the boundary at
$+0.38$~ERB, the value that best separates both human \class{hll} from \class{lhh} and the models'
own two conditions; it is above zero because the models' assertions are flat rather than falling
(\S\ref{app:deadband}).

The interventions of \S\ref{sec:where} are scored by three read-outs of the final-word contour,
each a signed number that is larger for a more question-like contour: the excursion; the
\emph{velocity} (the mean slope of a smoothed contour); and a \emph{cluster} read-out. To define
the latter, every contour is centred on its speaker's mean pitch, the unintervened recordings'
contours are split into two groups by $k$-means, and the read-out is defined as a contour's distance
to the falling group's centroid minus its distance to the rising group's centroid. For a read-out
$\Phi$, an intervention's \emph{normalized displacement} $\Delta_\Phi$ is the mean change it causes
(the mean taken over all intervened recordings) divided by the difference between the unintervened
conditions' means, so $0$ is no effect and $1$ moves a recording all the way to the other condition
(\S\ref{app:readouts}). We also apply a probe trained on human speech from HuBERT encodings to
classify the edge as rising or falling, reported with its accuracy on unmodified generations
(\S\ref{app:probes}). Finally, we use nearest-neighbor classification to describe each final-word
contour by one of the five tune shapes that emerge from the human productions (\S\ref{app:shapes}):
\class{lhh}; \class{lll}; the \emph{mixed edge} \class{lhl}/\class{llh} which is neither clearly falling
nor rising; \class{hll}/\class{hlh}; and the \emph{high plateau} \class{hhl}/\class{hhh}. Every
generated contour is centred on the mean and divided by the standard deviation of its own voice's
unsteered recordings, as each human contour is by its speaker's. We use this to describe shape in the corpus's terms; we are not claiming that
the model manipulates abstract tune representations.

\section{Localization via interchange interventions}
\label{sec:where}\label{sec:results}

In this section, we use interchange interventions \citep{geiger2021causal} to localize the
computation of the tune among the backbone transformer, the L0-head and the depth transformer,
and among the levels of the token stack, a causal counterpart to the probing results of
\citet{vigneaux2026probing}.

\paragraph{The tune is carried by the end of the final word.} Before intervening, we establish on
which part of the utterance the two conditions differ: the systematic pitch-contour variability
induced by the \class{A} or \class{I} condition is only expressed on the final word of the sentence
(\S\ref{sec:material}), and within that word at its offset and not at its onset
(Figure~\ref{fig:final-word}),  This fixes where an intervention must act and what it must move:
the edge of the tune rather than its pitch accent. (Note that expectations from tunes produced by humans are that falling and rising tunes will differ at both the onset and offset.)

\begin{figure}[t]
\centering
\includegraphics[width=0.94\linewidth]{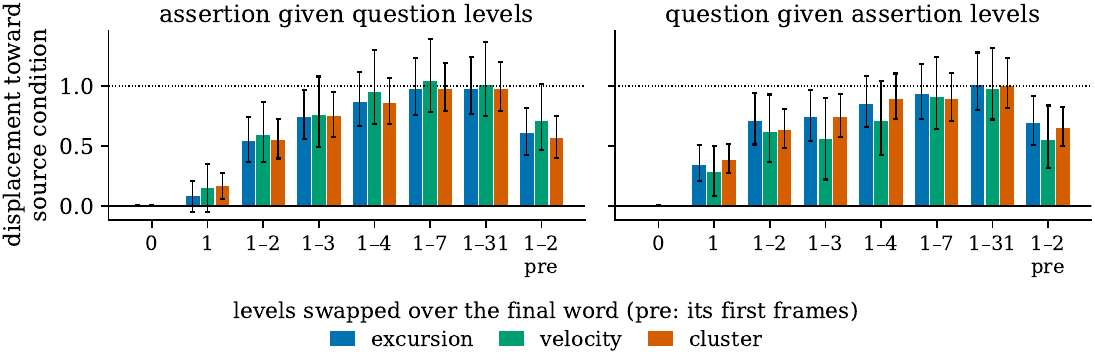}
\caption{Normalized displacement toward the source condition when levels of the token stack are
swapped over the final word between a matched assertion and question, by range of levels and by
read-out (CSM, ten-speaker set, $90$ pairs, paired-bootstrap intervals; $1$ means the swapped recording lands on
the source condition's mean). The last group swaps levels $1$--$2$ over the word's first frames
(\emph{pre}) instead of its last (\S\ref{app:swap}).}
\label{fig:swap}
\end{figure}

\begin{table}[t]
\centering\small
\setlength{\tabcolsep}{3.5pt}
\begin{tabular}{@{}l rcc rcc@{}}
\toprule
 & \multicolumn{3}{c}{CSM ($90$ pairs)} & \multicolumn{3}{c}{Fish S2 ($90$ pairs)} \\
\cmidrule(lr){2-4}\cmidrule(l){5-7}
Intervention & $\Delta_\Phi$ & rising, \class{A}/\class{I} & WER & $\Delta_\Phi$ & rising, \class{A}/\class{I} & WER \\
\midrule
Unmodified & --- & $17$/$93$ & $0.257$ & --- & $17$/$98$ & $0.130$ \\
Swap levels $1$--$K{-}1$ & $0.99$ {\scriptsize $[0.97, 1.01]$} & $90$/$18$ & $0.328$ & $0.91$ {\scriptsize $[0.82, 0.99]$} & $91$/$17$ & $0.194$ \\
Swap level $0$ & $0.00$ {\scriptsize $[0.00, 0.00]$} & $17$/$93$ & $0.266$ & $0.09$ {\scriptsize $[0.00, 0.21]$} & $18$/$92$ & $0.228$ \\
Re-run, own $\vh_t$ and $q^0_t$ & $0.01$ {\scriptsize $[-0.03, 0.03]$} & $21$/$92$ & $0.336$ & $0.00$ {\scriptsize $[-0.02, 0.02]$} & $13$/$96$ & $0.144$ \\
Source $\vh_t$, own $q^0_t$ & $0.98$ {\scriptsize $[0.94, 1.02]$} & $90$/$17$ & $0.348$ & $0.91$ {\scriptsize $[0.82, 0.98]$} & $90$/$19$ & $0.165$ \\
Own $\vh_t$, source $q^0_t$ & $0.01$ {\scriptsize $[-0.03, 0.04]$} & $20$/$92$ & $0.344$ & $0.11$ {\scriptsize $[0.06, 0.18]$} & $23$/$91$ & $0.185$ \\
Source $\vh_t$ and $q^0_t$ & $1.00$ {\scriptsize $[0.97, 1.03]$} & $90$/$19$ & $0.337$ & $0.99$ {\scriptsize $[0.97, 1.02]$} & $96$/$16$ & $0.172$ \\
Steer, probe direction$^{\ddagger}$ & $0.05$ {\scriptsize $[-0.03, 0.15]$} & $18$/$91$ & $0.254$ & $0.14$ {\scriptsize $[0.03, 0.28]$} & $21$/$91$ & $0.126$ \\
Steer, mean difference$^{\S}$ & $0.74$ {\scriptsize $[0.28, 1.12]$} & $39$/$38$ & $0.305$ & $0.76$ {\scriptsize $[0.46, 0.96]$} & $46$/$29$ & $0.122$ \\
\bottomrule
\end{tabular}

\caption{Interventions in both systems. $\Delta_\Phi$: normalized displacement toward the source
condition, mean over the three read-outs and both directions, with the range over those six cells.
\emph{Rising}: per cent of assertion (\class{A}) and question (\class{I}) targets whose final-word
excursion exceeds $+0.38$~ERB afterwards. WER: word error rate. Middle blocks: the depth model
re-run with $\vh_t$ and $q^0_t$ from the target (own) or the source. Steering rows: the offline edit
of \eqref{eq:gap} at $\alpha = 4$ gaps; $^{\ddagger}$the probe direction's most effective strength.}
\label{tab:interventions}
\end{table}

\paragraph{Token swaps: the decoder's sensitivity.}
\label{sec:swap}\label{sec:truncation}
An interchange intervention modifies one recording, the \emph{target}, with values taken from
the other recording of its matched pair, the \emph{source}. Swapping the sampled tokens at every
level the depth model emits (the full swap, levels $1$ to $K{-}1$) \emph{over the final word}
between a matched assertion and question (i.e., same text and speaker, different tune), followed by
codec decoding, displaces contours from one condition to another under all three read-outs (Figure~\ref{fig:swap},
Table~\ref{tab:interventions}). Level $1$ alone does little in CSM ($\Delta_\Phi = 0.23$); levels
$1$--$3$ move the recording about three quarters of the way, and levels $1$--$7$ as far as the full
swap. Swapping level $0$ changes neither the tune nor the words; a more detailed analysis with Mimi
shows that the final word's log-mel spectrogram moves by $1.7$~dB on average, against $11$~dB for
the full swap (\S\ref{app:level0-audio}); what level $0$ carries into the other levels is tested by
the depth-model re-runs below. Word error rate rises slightly for the full swap and more for
partial ranges. Decoding truncated stacks (the first $k+1$ levels only), the contrast between conditions appears at $k = 1$ and reaches the
separation of the full stack at $k = 2$, both in terms of excursion and clustering
(\S\ref{app:depth}). S2 reproduces this with a different codec and language model; with
ten levels instead of thirty-two, its level $1$ alone already carries most of the contrast, and
its level $0$ a small part of it (\S\ref{app:s2}).

\paragraph{The backbone state is the route; level $0$ is not.}
\label{sec:backbone}\label{sec:backbone-patch}
Treating the generator as a causal model over its activations and substituting one variable at a
time \citep{geiger2021causal}, we re-run the depth model $p_\psi$ on a target frame with either of its
two inputs replaced by the source's. Giving it the source's $\vh_t$ flips the tune as completely as
the full swap: the final word's offset moves to the source condition while its onset is
unchanged. Giving it the source's level-$0$ tokens, which differ from the target's in most frames,
changes most of the depth model's sampled tokens, but not the tune: the excursion does not move in CSM,
and moves by about a tenth of the contrast in S2 (Table~\ref{tab:interventions}). The words need
both inputs. Across two different sentences, the depth model given one sentence's $\vh_t$ and the
other's level-$0$ tokens garbles the final word, and produces the other sentence's word only when
given both of its inputs (\S\ref{app:level0}). The route is therefore
\begin{equation}
\text{prompt} \xrightarrow{f_\theta} \vh_t \xrightarrow{p_\psi} q^{1:K-1}_t \to \text{tune},
\quad
\text{prompt} \xrightarrow{f_\theta,\,p_\phi} (\vh_t, q^0_t) \xrightarrow{p_\psi} q^{1:K-1}_t \to \text{words}.
\label{eq:route}
\end{equation}
The factorization \eqref{eq:staged} lets $q^0_t$ influence everything $p_\psi$ emits; that it
carries the words but not the tune is a property of the trained models.

\paragraph{A mean difference steers it, offline.}
\label{sec:steering}\label{sec:s2-route}
Write $\vx_t$ for the backbone's final-layer state before its normalization, so that
$\vh_t = \mathrm{norm}(\vx_t)$. Let $\bar\vx_{\class{I}}$ and $\bar\vx_{\class{A}}$ be the mean states over
the final words of each condition, and
\begin{equation}
\vw = \bar\vx_{\class{I}} - \bar\vx_{\class{A}}, \qquad
g = \lVert \vw \rVert \;\;(\text{the \emph{gap}}), \qquad
\hvw = \vw / g .
\label{eq:gap}
\end{equation}
Replacing the state over the final word by $\mathrm{norm}(\vx_t + \alpha g \hvw)$, so that
$\alpha$ is measured in gaps, and re-running the depth model moves the tune toward the other
condition in both systems, about three quarters of the way at four gaps
(see Table~\ref{tab:interventions} and \S\ref{app:offline}). A logistic-regression probe fitted to the
same states separates the conditions as well, but its direction, used the same way, barely moves
the tune at any strength, although the two directions are correlated (cosine $0.64$ to $0.68$ in
CSM, $0.50$ to $0.55$ in S2), and its component orthogonal to $\hvw$ does nothing on its own
(\S\ref{app:offline}). We steer with the difference of means.

\paragraph{Control, but not the whole story.}
\label{sec:offline-limits}
The offline edit controls the tune, with two limits.
\begin{itemize}\itemsep2pt
\item \emph{The edit is not symmetric, and one direction damages the voice.} The edit brings the
final word's $F_0$ offset for either condition onto the other's natural range, but not at the same
price (Figures~\ref{fig:offline} and~\ref{fig:offline-s2}). Assertions pushed toward a question keep their voicing and
harmonicity: the push mostly raises their pitch, and their $F_0$ trajectories most closely resemble the
high plateau \class{hhl}/\class{hhh}.
Questions pushed toward an assertion steadily degrade in voicing and harmonicity once the offset
approaches the target, whether the edit is made before or after the backbone's normalization and
whether or not level $0$ is redrawn from the edited state (\S\ref{app:offline}).
\item \emph{The vector might just encode the punctuation mark.} The two conditions differ in one
character of the prompt (the punctuation), so $\hvw$ could encode the orthographic sign alone, leaving the depth
model to turn it into a tune; the intervention would then be an expensive way of writing a
question mark. It is not: $\hvw$ is aligned with the rising-versus-falling direction fitted on
human recordings whose transcripts carry no punctuation (\S\ref{sec:human}; cosine $0.66$ in CSM,
$0.44$ in S2), far beyond a shuffled-label null (\S\ref{app:directions}). The
state of a punctuated prompt therefore already represents intonation. It still encodes the whole
prompt as well, and we conjecture that this contributes to the damage; a direction fitted on tune
differences alone does steer more cleanly (\S\ref{sec:human}).
\end{itemize}

\section{Steering with tunes taken from human speech}
\label{sec:human}

\paragraph{Steering vectors derived from human imitations.} The corpus gives, for the same three sentences and the
same ten speakers, productions of all eight tunes. We feed each recording to the generator as a
voice prompt is fed --- its transcript, which carries \emph{no punctuation}, then its audio codec tokens
--- and record the backbone's final-layer state before normalization, $\vx_t$, at every frame. A
difference of means over the frames of the final word, between the recordings of two tunes, gives
a unit vector $\hvw$ and a gap $g$ as in \eqref{eq:gap}. The \textsf{tune} vector contrasts
\class{lhh} with \class{hll}, the paradigmatic tunes of interrogation and assertion, so it pushes
toward a lower pitch accent and a higher edge (\S\ref{app:targets}). The two groups contain the
same speakers reading the same words, so nothing in the input distinguishes them but the tune. Vectors fitted on
the edge tones alone (\class{xhh} against \class{xll}, with \class{x} for either tone), where \S\ref{sec:where} located the
conditions' difference, and on the pitch accent alone are reported in \S\ref{app:edge} and taken up
in \S\ref{sec:discussion}. Here we keep the contrast over the holistic tune, as our current focus.
For space, the figures of this section show CSM; S2's, which are as informative, are
Figures~\ref{fig:offline-s2} (offline) and~\ref{fig:alpha-s2} (in the loop).

\paragraph{Steering procedure.} The edit acts inside the generation loop. An \class{A} or
\class{I} recording is generated and aligned, which locates the onset $t^*$ of its final word; the
generation is then replayed, drawing the same tokens up to $t^*$ and, from $t^*$, replacing the
state before normalization by $\vx_t + \alpha g \hvw$ for as many frames as the source word
occupied. Every level is sampled under the edit, and the model decides its own timing and when to
stop. Strength $\alpha$ is in gaps, so one grid ($1.5$ to $4.5$) is comparable across vectors. At
$\alpha = 0$ the replay reproduces its source exactly in CSM (\S\ref{app:inloop}).

\begin{figure}[t]
\centering
\includegraphics[width=0.96\linewidth]{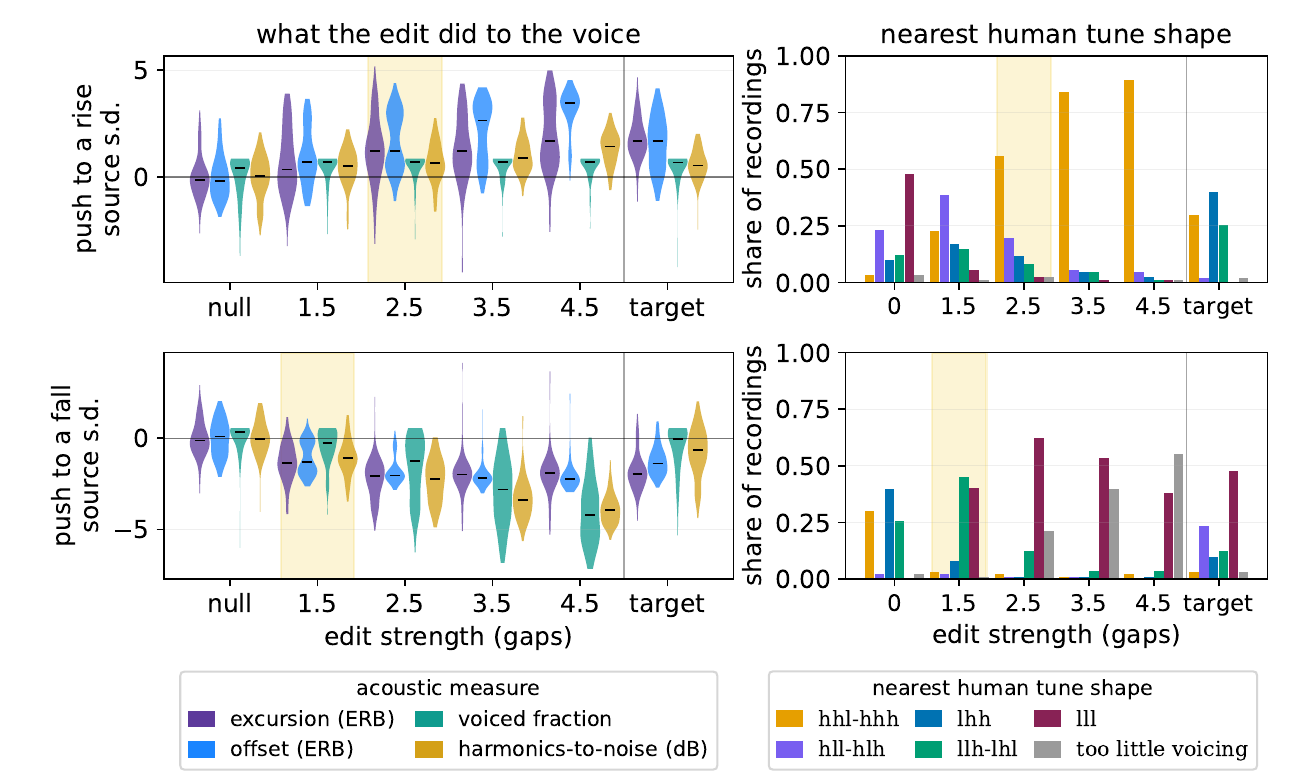}
\caption{Steering along the \textsf{tune} vector in the loop, CSM, $90$ recordings per strength:
assertions pushed toward a rise (top), questions toward a fall (bottom). \emph{Left:} acoustic
measures over the final word in units of the source condition's s.d., centred on the unsteered run
(\emph{null}); rightmost, the unmodified target condition, so a strength is read as a position
between source and target. \emph{Right:} nearest human tune shape (\S\ref{sec:measures}), with the
target condition's own distribution at the right; grey, too little voicing (\S\ref{app:shapes}). Shaded: the strength
chosen. S2: Figure~\ref{fig:alpha-s2}; offline: Figure~\ref{fig:offline} (CSM) and \ref{fig:offline-s2}  (S2).}
\label{fig:alpha}
\end{figure}

The strength is chosen for each system and direction as the one at which the steered recordings'
distributions of excursion, offset, voiced fraction and HNR come closest to those of the target
condition's unmodified recordings: the mean over the four measures of the Wasserstein distance, in
units of the unsteered runs' standard deviation (\S\ref{app:strength}). It selects $\alpha = 2.5$
gaps for the push toward a rise and $\alpha = 1.5$ for the push toward a fall, in each system
separately (Figures~\ref{fig:alpha} and~\ref{fig:alpha-s2}), and the same strengths in each of ten
folds when it is applied leaving one speaker out. Beyond it, a push toward a rise makes the voice
artificially high, and a push toward a fall collapses voicing and HNR. A recording \emph{flips} when its final-word
excursion crosses to the other condition's side of $+0.38$~ERB, the boundary that best separates the
two conditions (\S\ref{sec:measures}). A flip is thus a move to the other condition, not necessarily
to its paradigmatic tune: since the models' own assertions have a  flat rather than falling pitch trajectory, a question
that flips toward an assertion may end flat.

\paragraph{What the edit costs the voice.} Over every steered recording at the strengths we read,
flipped or not, UTMOS stays within $0.2$ of the target condition's own
recordings in both systems and both directions, and speaker similarity to the voice prompt is
unchanged (\S\ref{app:quality}). The final word's harmonicity separates the two directions: pushed
toward a rise it matches or exceeds that of the model's own questions, and pushed toward a fall
it is about $1.5$~dB below that of its assertions in CSM and $2$~dB in S2; stronger edits flip
more recordings toward a fall only at a steep cost in harmonicity (Figure~\ref{fig:flip-quality}).
The asymmetry of \S\ref{sec:offline-limits} is thus much reduced inside the loop, but not gone. The
median change in word error rate against each recording's unsteered run is zero in every cell.

\begin{figure}[t]
\centering
\includegraphics[width=\linewidth]{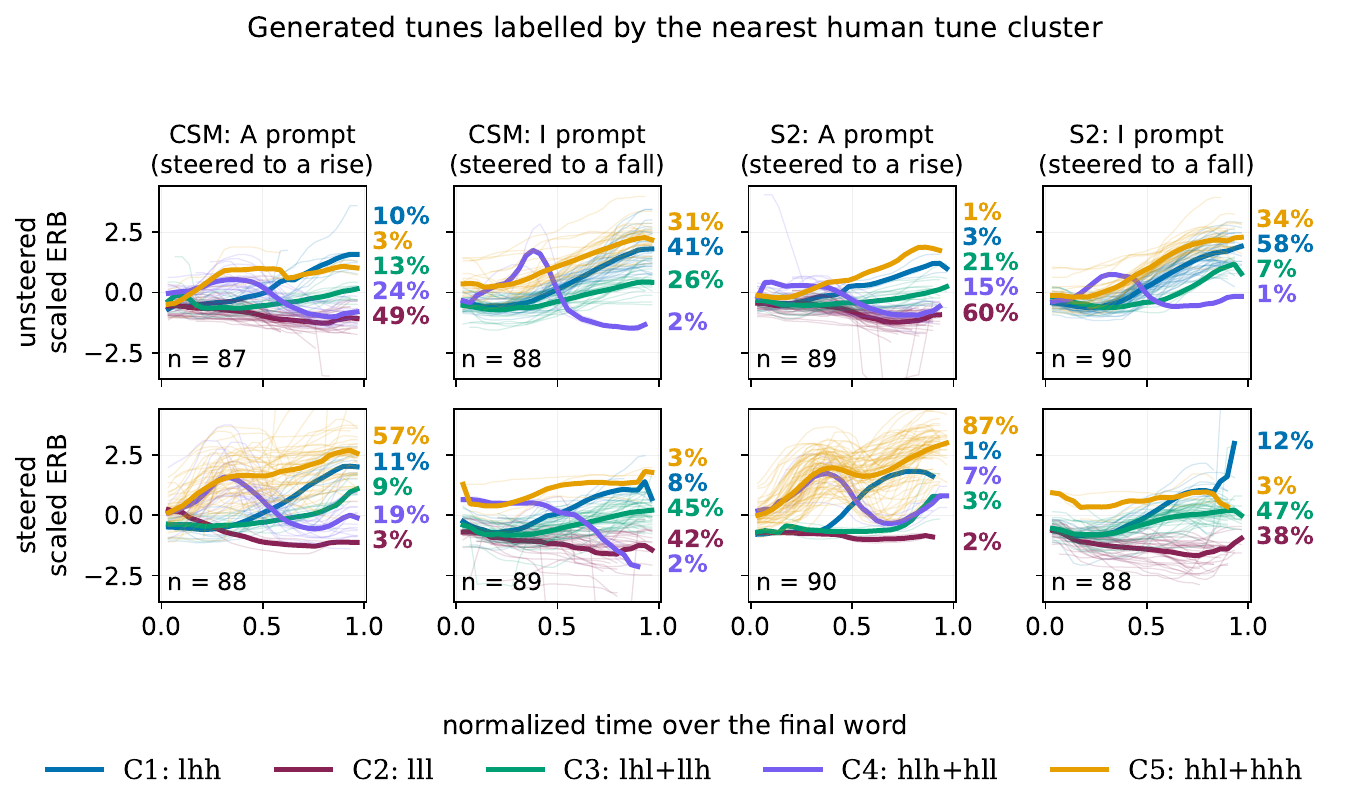}
\caption{$F_0$ trajectories of every generated recording, coloured by the nearest of the five
human tune clusters (Figure~\ref{fig:cole}), in scaled ERB, each voice scaled by its own unsteered
recordings as the clusters were scaled by speaker; nothing is fitted to the generated material. Bold: the mean of each colour, with its share of the panel; words with too little voicing
are left out. Columns: assertions steered toward a rise, questions toward a fall; rows: unsteered,
steered.}
\label{fig:centroids}
\end{figure}

\paragraph{Where the steered recordings land.} Figure~\ref{fig:centroids} labels every generated
recording by the nearest of the corpus's five tune shapes. The unsteered prompts produce the
paradigmatic tunes only in part: the models' assertions are mostly \class{lll} ($49\%$ in CSM,
$60\%$ in S2) rather than the predicted \class{hll}/\class{hlh} ($24\%$, $15\%$), and their
questions are the expected \class{lhh} in $41\%$ and $58\%$ of recordings, with most of the rest
on the high plateau \class{hhl}/\class{hhh} ($31\%$, $34\%$) or the mixed edge
\class{lhl}/\class{llh} ($26\%$, $7\%$). Steering moves the recordings out of their unedited (target's) shape
in every case. Pushed toward a rise, they mostly
reach the high plateau ($57\%$, $87\%$) and rarely \class{lhh} ($11\%$, $1\%$); pushed toward a
fall, they divide between the mixed edge \class{lhl}/\class{llh} ($45\%$, $47\%$) and
\class{lll} ($42\%$, $38\%$), and almost never reach \class{hll}/\class{hlh}
($2\%$, $0\%$). Although the \textsf{tune} vector also contrasts a low with a high pitch accent,
the effect of steering is systematic only on the edge tone. The edit reliably takes a recording off its unedited
tune, and less reliably delivers it to the intended one.

The shapes it delivers are, however, ones that neither punctuation mark produces often. The high
plateau makes up $57\%$ and $87\%$ of the steered rises, against $31\%$ and $34\%$ of the models'
own questions; the mixed edge $45\%$ and $47\%$ of the steered falls, against $13\%$ and $21\%$
of their own assertions. The intervention thus widens the set of tunes beyond what the prompt
selects, although it does not yet choose among them.

\paragraph{Held-out speakers.}\label{sec:loso} Refitting the \textsf{tune} vector on nine speakers
and steering the tenth, for each speaker in turn, flips $60\%$ of CSM's recordings in both
directions and $76\%$ and $58\%$ of S2's, against $56\%$, $56\%$, $72\%$ and $57\%$ of the original
intervention, with the same voice quality (\S\ref{app:directions}).

\paragraph{Beyond punctuation.} We run a pilot to test the possibility of obtaining a tune that no
punctuation mark elicits: a vector contrasting \class{hlh} (a fall-rise) with \class{lll}, fitted as
the other imitation-derived vectors are, was applied to the $43$ CSM assertions whose unsteered run
is nearest \class{lll}. Each contour is labelled by the nearest of eight per-tune templates, since
the five clusters merge \class{hlh} with \class{hll} (\S\ref{sec:hlh}, Figure~\ref{fig:hlh}).
Unsteered, $98\%$ are nearest \class{lll} and none \class{hlh}. At $2.5$ gaps, $26\%$ are nearest
\class{hlh}, the most of any template (\class{lll} also $26\%$); from $3.5$ gaps they move to
\class{hhl} ($56\%$ to $75\%$), keeping the high accent and losing the boundary. At the right
strength, a vector fitted on human imitations of a tune without an orthographic cue thus produces
that tune in a quarter of the recordings.

\section{Discussion}
\label{sec:discussion}

In both systems the pitch contour is written by the depth model from a backbone state that already
encodes it (at least partially). The level-$0$ tokens, computed separately from the backbone, carry
little to no intonation, and in CSM the waveform hardly depends on them at all. To control or
fine-tune intonation in a staged generator, the lever is the backbone state, not the `semantic'
level-$0$ audio token; more broadly, localization results as in \S\ref{sec:where} might guide
fine-tuning with prosodically annotated data, which is relatively scarce. Control also works best inside
the full generation loop: applied ``offline'', where the number of frames is fixed and changes in
audio tokens do not propagate in time, the vector moves the tune at the cost of the voice; inside
the loop the cost is small, because the model is free to choose the codewords, timing and length
that go with the tune it is pushed toward.

The alignment of the imitation-derived and punctuation vectors says that the state of a punctuated
prompt already represents intonation: the orthographic cue has been translated into something the
model also computes when it reads the pitch of human speech. It does not show that the model
operates with the tune categories of the autosegmental-metrical analysis. We use those categories
as units of analysis, to describe the shapes a model produces and how they depart from the
prototypes speakers produce, and the departures are informative: the models' assertions are flat
\class{lll} more often than the falling \class{hll} their prompt would predict, and their questions
often end on a high plateau. S2 departs further (\S\ref{app:edge}): a vector fitted on the pitch
accent alone flips up to $74\%$ of its questions toward an assertion ($34\%$ in CSM), which it can
only do through the edge, since the conditions do not differ where the accent sits. S2 thus seems
to represent pitch accent and edge tone jointly, although they vary independently in human speech.

For the contrast studied here, punctuation already does what steering does; the interest of
steering is that it acts on the representation the prompt produces, and so may reach tunes that no
orthographic cue elicits. It already takes recordings to shapes that neither punctuation mark
produces often, and a vector fitted for one such tune, \class{hlh}, reaches it in a quarter of the
recordings at the right strength. Choosing a tune reliably, across strengths and for the other
tunes of the corpus, and determining if imitation-derived vectors compose in linguistically
predictable ways, are challenges that remain open.

\section{Limitations}
\label{sec:limitations}
\label{sec:remaining}

The material is English imitated speech, three carrier sentences, ten speakers, and the tunes of a
single corpus. The voice prompt is as varied as it can be in pitch movements,
with all eight tunes, but each of them occurs over a single word, at the end of its recording. The steering
strengths were chosen on the ten-speaker set; the choice does not change when each speaker is held
out in turn, and the vectors were also held out
(\S\ref{sec:loso}). No listening test was run: voice quality is measured acoustically and by
prediction (harmonicity, voicing, UTMOS, speaker similarity), and tune shapes are read from $F_0$
contours, so a word with too little voicing cannot be labelled.\label{sec:endofbody}

\section*{Reproducibility statement}
\label{sec:endofmaintext}

 Code
accompanies the submission (first page). Every measurement reported here is produced by a script that reads a committed CSV; no number is
transcribed by hand. Figure regeneration runs from two entry points: the offline figures and tables
from the repository's figure and table scripts, and the in-loop figures --- the strength-choice
panels, the nearest-centroid labels and the reproduction of \citet{cole2023} --- from the steering
matrix's own scripts, which are released with it. The acoustic front end, the alignment policy,
the swap operator, the depth re-run and the effect-size estimator are covered by $354$ tests. The
offline depth re-runs are additionally checked against each generator's own computation
(\S\ref{app:rerun}) before any result computed from them is reported, and every in-loop run is
replayed at $\alpha = 0$ against its original recording (\S\ref{app:inloop}). Split seeds,
sampling temperature, top-$k$ and the per-recording generation seeds are stored inside each token
archive, as is the backbone state $\vh_t$ for every frame. Strength is quoted throughout in units of
each vector's own gap $g$ \eqref{eq:gap}; the in-loop edit is applied before the normalization,
over the aligned final word; and the rise/fall boundary is $+0.38$~ERB (\S\ref{app:deadband}).

\section*{AI statement}

In this work, we used generative AI tools to assist us in refining hypotheses, designing or
providing feedback on research methodology (particularly statistical methods), designing or providing feedback on experiments,  implementing
methods as code, and  interpreting results. We did not use generative AI tools to help develop theoretical models or
conceptual frameworks, formulate mathematical claims, clean or reformat a dataset, or support
qualitative or thematic data analysis. The remaining tasks listed in the
disclosure policy (generating synthetic datasets, 
assisting with translation, providing critical ingredients for proving mathematical claims, and
assisting in the writing of proofs) do not apply to this work. We also used generative AI tools to
assist with creating scientific figures and tables, creating and editing software code, drafting
parts of the paper, summarizing and analyzing existing literature, identifying relevant literature,
formatting references, and brainstorming.

The paper was not generated automatically: it was developed over multiple iterations across
several months, and every methodological decision was made deliberately and discussed among the
coauthors. We have reviewed all AI-assisted work; in particular, one of the authors tested and
reviewed the code, and we verified the authenticity and relevance of the references. We take
responsibility for the final content of this work, including any text, claims or artifacts
produced with the aid of generative AI.

\bibliographystyle{iclr2027_conference}
\bibliography{references}

\appendix
\FloatBarrier
\section{Background}
\label{app:background}

\FloatBarrier
\subsection{Residual quantization}
\label{app:rvq}

Let $\enc : \Xset \to \R^{D \times T}$ be a codec encoder mapping a waveform to $T$ latent frames
of dimension $D$, and let $\Cb_k = \{e^1_k, \dots, e^V_k\} \subset \R^D$ be the codebook at level
$k$, for $k = 0, \dots, K-1$. Residual vector quantization of a frame $z \in \R^D$ proceeds
greedily down a stack of levels: setting $r_{a} = z$ at the first level $a$ of the stack,
\begin{equation}
  q_k \;=\; \arg\min_{v \in \{1,\dots,V\}} \bigl\| r_k - e^v_k \bigr\|_2 ,
  \qquad
  r_{k+1} \;=\; r_k - e^{q_k}_k ,
  \label{eq:rvq}
\end{equation}
so that each level quantizes what the levels before it left unexplained. Both codecs we study
quantize level $0$ on its own, with a plain vector quantizer whose codeword is trained to predict
the representation of a self-supervised speech encoder (\S\ref{sec:setup}), and differ in what
the other levels quantize. In S2's codec, levels $1$ to $9$ are a residual stack \eqref{eq:rvq}
started at $r_1 = z - e^{q_0}_0$, the part of the frame level $0$ leaves
unexplained (in its released codec code; \citealp{fishaudio2026s2}). Mimi instead uses a \emph{split} quantizer: levels $1$ to
$31$ are a residual stack started at $r_1 = z$, in parallel with level $0$, so that the acoustic
levels are not constrained to encode only what level $0$ misses \citep{defossez2024moshi}; Moshi runs
it with $8$ levels, CSM with $32$. In both,
the decoder reads the sum of the selected codewords, and the \emph{partial reconstruction} at
depth $k$ is the waveform decoded from the first $k+1$ levels,
\begin{equation}
\hat z^{(k)} \;=\; \sum_{j=0}^{k} e^{\,q^j}_j,
\qquad
\hat x^{(k)} = \dec\big(\hat z^{(k)}\big).
\label{eq:partial}
\end{equation}
Within the residual stack the greedy rule orders the levels from coarse to fine, which is what the
swap of \eqref{eq:swap} and the truncation of \S\ref{app:depth} rely on. In Mimi, level $0$ and
level $1$ each quantize the whole frame, so a level-$0$ swap exchanges a code that the acoustic
levels do not depend on, and a truncation to levels $0$--$1$ keeps the coarsest acoustic
description of the frame in addition to the semantic one.

\FloatBarrier
\subsection{Tune notation}
\label{app:tunepair}\label{app:notation}

Following the autosegmental-metrical convention \citep{pierrehumbert1980}, a nuclear tune is
written as three tone symbols, pitch accent, phrase accent and boundary tone, so \class{hlh} is a
high pitch accent, a low phrase accent and a high boundary tone; the eight monotonal tunes (one
tone per position) are the cross product of two levels at each position. An \class{x} stands for
either tone, so \class{xhh} is every tune ending in a high phrase accent and a high boundary tone.

\FloatBarrier
\FloatBarrier
\section{Measurement}
\label{app:measurement}

\FloatBarrier
\subsection{The acoustic front end}
\label{app:frontend}

$F_0$ is tracked with Praat's filtered-autocorrelation method \citep{boersma2024praat} and
converted to the equivalent-rectangular-bandwidth scale
\begin{equation}
  \E(f) \;=\; 21.4 \, \log_{10}\!\bigl(1 + f/228.8\bigr),
\end{equation}
on which a difference means the same size of pitch movement whether the speaker is high or low in
their range \citep{hermes1991erb}. The contour is restricted to the final-word interval given by
forced alignment and resampled to $N = 30$ points. Grid points outside the voiced extent are left
undefined rather than extrapolated, so unvoiced margins bias neither an average nor the position of
a peak, and nothing is imputed. Short voiced islands at the edge of the word, detached from the
main voiced body by an unvoiced gap, are marked unvoiced before resampling: they are cross-word
pitch leaking past an imperfect alignment boundary, and interpolating across the gap fabricates a
steep slope. An island is removed only if it is shorter than a tenth of the word and lies within a
tenth of an edge.

The \emph{excursion} of a contour $c$, with $\mathcal{V}$ its voiced points, is the level of its
last fifth minus that of its first:
\begin{equation}
\mathrm{exc}(c) \;=\;
\frac{1}{|\mathcal{V}_{\text{last}}|}\sum_{i \in \mathcal{V}_{\text{last}}} \E(c_i)
\;-\;
\frac{1}{|\mathcal{V}_{\text{first}}|}\sum_{i \in \mathcal{V}_{\text{first}}} \E(c_i),
\qquad
\mathcal{V}_{\text{last}} = \mathcal{V} \cap \{\lceil 4N/5 \rceil, \dots, N\},
\label{eq:excursion}
\end{equation}
and the \emph{onset} and \emph{offset} are its two terms. The \emph{velocity} is the derivative of
a cubic regression spline with four interior knots, about a fifth of the word apart ---
roughly $110$~ms for a median word, the timescale of pitch accents and boundary tones --- taken
analytically; a linear interpolation differentiated numerically turns pitch-frame steps into a
zigzag. The HNR uses Praat's cross-correlation method with a $10$~ms step and a $75$~Hz floor,
averaged over voiced frames. The cluster read-out centres each contour $F$ of speaker $v$ on
that speaker's mean over the $n_v$ unmodified recordings,
\begin{equation}
  F \;\mapsto\; F - \tfrac{1}{n_v N}\textstyle\sum_{i \in v,\, j} F_{ij} ,
\end{equation}
which keeps pitch level relative to the speaker's register; the mean is taken over voiced points,
and an intervened recording is centred on the mean of its target's speaker.

\FloatBarrier
\subsection{The rise/fall boundary}
\label{app:deadband}

A recording is counted as rising or falling by which side of $+0.38$~ERB its excursion falls on.
The boundary is shifted off zero because the models' assertions are flat rather than falling ---
their mean excursion is $-0.01$~ERB, whereas, for comparison, human imitations of \class{hll} fall
on average by $1.04$~ERB and human imitations of \class{lll} by $0.72$~ERB --- so that a criterion
taken at zero scatters them across both sides: on the ten-speaker set it reads $27\%$ of CSM's
unsteered assertions as rises against $17\%$ at $+0.38$. The value is not chosen to suit the
models: the boundary that best separates the human imitations of \class{hll} and \class{lhh} is
$+0.375$~ERB, and the one that best separates the models' own two prompt conditions is $+0.381$.

\FloatBarrier
\subsection{The read-outs and the normalized displacement}
\label{app:readouts}

The three read-outs of \S\ref{sec:measures} each give a signed score per recording: the
excursion, the mean velocity, and, for the cluster read-out, a contour's distance to the
falling-like centroid minus its distance to the rising-like one, of two $k$-means clusters fitted to
the unmodified recordings' speaker-centred contours. Which cluster is rising-like is decided by its members' mean
excursion, never by the design labels. For a read-out $\Phi$, unmodified target recordings
$x_1, \dots, x_n$, their intervened counterparts $\swapop x_i$, and the mean score $\bar\Phi_c$ of
the unmodified recordings in condition $c$,
\begin{equation}
  \Delta_\Phi
  \;=\;
  \frac{\frac{1}{n}\sum_{i=1}^{n}\bigl[\Phi(\swapop x_i) - \Phi(x_i)\bigr]}
       {\bar\Phi_{\text{source}} - \bar\Phi_{\text{target}}} ,
  \label{eq:displacement}
\end{equation}
where $i$ indexes recordings, not frames. The numerator is the mean within-pair movement and the
denominator the separation between the conditions on unmodified recordings, so $\Delta_\Phi$ is
dimensionless and comparable across read-outs. Intervals are paired bootstraps over pairs.

Table~\ref{tab:families} scores the three read-outs against the design labels on the unsteered
ten-speaker sets: each recovers the conditions for $84$ to $94$ per cent of recordings. The
cluster read-out centres each contour on its speaker's mean, as \citet{cole2023} scale contours by
speaker, so that ten voices whose registers differ by more than the tune contrast are not split by
voice.

\begin{table}[!htbp]
\centering\small
\begin{tabular}{l ccc ccc}
\toprule
& \multicolumn{3}{c}{CSM ($180$)} & \multicolumn{3}{c}{S2 ($180$)} \\
\cmidrule(lr){2-4}\cmidrule(l){5-7}
read-out & agreement & AUC & ARI & agreement & AUC & ARI \\
\midrule
excursion              & $0.84$ & $0.90$ & $0.46$ & $0.84$ & $0.96$ & $0.47$ \\
velocity               & $0.84$ & $0.88$ & $0.46$ & $0.84$ & $0.96$ & $0.47$ \\
cluster                & $0.84$ & $0.90$ & $0.46$ & $0.94$ & $0.97$ & $0.77$ \\
\bottomrule
\end{tabular}
\caption{The three read-outs on the unsteered ten-speaker sets, scored against the design labels
with the score's sign at zero. \emph{Agreement}: fraction of recordings put in the right condition; AUC: area under the ROC
curve; ARI: adjusted Rand index of the partition.}
\label{tab:families}
\end{table}

\FloatBarrier
\subsection{How much the measurement depends on the alignment}
\label{app:robustness}

Swapped recordings are not force-aligned: a swap leaves level $0$ untouched and does not change the
frame count, so their segmental timing should match the target recording's, and the target's
final-word interval is inherited. What can be checked is whether the measurement is sensitive to
the boundary at all. On $84$ aligned CSM recordings of an earlier pilot set, prompted with one
recorded reference voice, we recomputed the excursion under boundaries
displaced by $\pm 40$ and $\pm 80$~ms --- one Mimi frame is $80$~ms --- and under no alignment at
all, taking the final two fifths of the utterance's voiced extent (Table~\ref{tab:robustness}).
Displacing the boundary by half a frame moves the excursion by $0.14$ to $0.23$~ERB against a
condition separation of $2.4$, and preserves its sign on $94$ to $99$ per cent of recordings. The
error is one-sided: displacing the window $80$~ms earlier shrinks the separation to $1.56$~ERB,
because the window then stops short of the final rise, while displacing it later costs almost
nothing. A swap that somehow shortened the recording would therefore bias the measured effect
down, not up.

\begin{table}[!htbp]
\centering\small
\begin{tabular}{lrrrr}
\toprule
Final-word interval & mean $|\Delta|$ & max $|\Delta|$ & $r$ & sign agreement \\
\midrule
displaced $-80$~ms                  & $0.482$ & $1.662$ & $0.930$ & $0.893$ \\
displaced $-40$~ms                  & $0.233$ & $1.191$ & $0.981$ & $0.940$ \\
displaced $+40$~ms                  & $0.140$ & $1.082$ & $0.992$ & $0.988$ \\
displaced $+80$~ms                  & $0.182$ & $1.569$ & $0.985$ & $0.976$ \\
no alignment (asserted interval)    & $0.218$ & $1.555$ & $0.980$ & $0.988$ \\
\bottomrule
\end{tabular}
\caption{Excursion under perturbed and absent final-word boundaries, against the excursion under
the forced alignment, in ERB. The condition separation is $2.409$~ERB under the forced alignment,
$2.046$ and $2.565$ under $\mp 40$~ms, $1.560$ and $2.565$ under $\mp 80$~ms, and $2.555$ with no
alignment.}
\label{tab:robustness}
\end{table}

\FloatBarrier
\subsection{The probes}
\label{app:probes}

Three probes trained on human speech were applied to the generated audio: a rising-against-falling
edge probe (\class{xll} against \class{xhh}), a pitch-accent probe (\class{hxx} against
\class{lxx}) and a five-class tune probe. All read HuBERT-base hidden states with attention pooling
over the final word, and each is the average of five random-split runs rather than the best of
them; held-out accuracies on human speech are $0.972$, $0.923$ and $0.819$. Applied to all $180$
unsteered recordings of each system, the edge probe recovers the generating condition at accuracy
$0.87$ in CSM and $0.92$ in S2 (area under the curve $0.89$ and $0.97$), against $0.88$ and $0.93$
for the excursion's sign at $+0.38$~ERB. The five-class probe is compared with the
nearest-centroid labels the paper reports in \S\ref{app:shapes}. The accent probe is at chance on unmodified
generations and is not used. This need not mean the probe is broken: the generated assertions and
questions share their pitch accent (Figure~\ref{fig:final-word}), so there is little accent
contrast in the generated material for a probe to recover, unlike in the human recordings it was
trained on. We have no working measurement of the pitch accent in generated audio, and make no
claim about it beyond the contour shapes of \S\ref{app:shapes}.

\FloatBarrier
\section{Prompt-side control}
\label{app:prompting}\label{app:controls}\label{app:tags}

\FloatBarrier
\subsection{Style tags}

Fish S2 accepts free-form inline instructions in square brackets, learned from captions generated
by a separate audio captioning model \citep{fishaudio2026s2}. Eighteen tags were chosen in three groups: credence
(\class{[completely convinced]}, \class{[confident]}, \class{[fairly certain]}, \class{[hedging,
not quite sure]}, \class{[very uncertain]}, \class{[doubtful]}, \class{[questioning]}), attitude
(\class{[sarcastic]}, \class{[surprised]}, \class{[skeptical]}, \class{[incredulous]},
\class{[impressed]}) and speech act (\class{[asking a question]}, \class{[stating a fact]},
\class{[calling out to someone across a distance]}, \class{[correcting a mistake]},
\class{[listing, with more items to come]}, \class{[confirming what was just said]}). Each
prefixes the three carrier sentences, which end in a full stop and carry an \class{[emphasis]}
marker on the sentence-final name so that the nucleus falls where it is measured, as does an
untagged control. Every voice of the ten-speaker set generates one example of each tag on each of
the three carrier sentences. Each voice is elicited with the same randomly ordered eight-tune voice
prompt as in the main experiments. Therefore, in total, we generate $570$ utterances ($19$ tag conditions, counting the untagged
control, $\times$ $3$ target sentences $\times$ $10$ speakers), which are force-aligned and scored by the
excursion \eqref{eq:excursion}.

\begin{figure}[!htbp]
\centering
\includegraphics[width=0.8\linewidth]{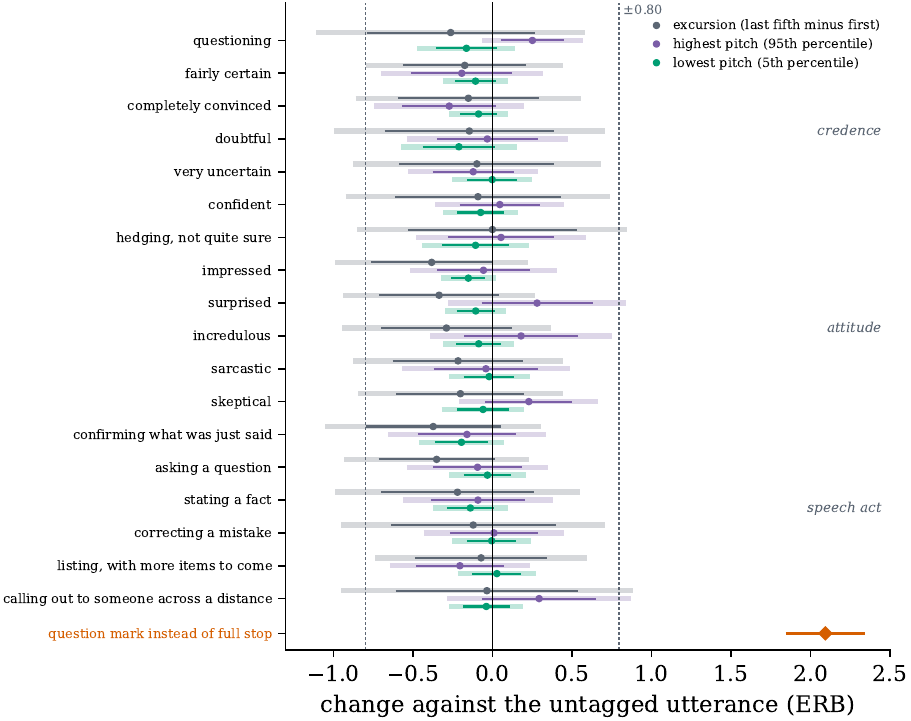}
\caption{Each style tag's mean change against the untagged utterance of the same voice and
sentence, over the ten speakers, in three measures of the final word: its excursion (grey), its
highest pitch (purple) and its lowest pitch (green), taken as the $95$th and $5$th percentiles of
its voiced contour so that one stray pitch point cannot set them. Each has its $95\%$ interval
(dark) and the Bonferroni interval over the eighteen tags (light); the dashed lines mark the
widest $95\%$ interval of the excursion. Below, the question mark's effect on the excursion in the
same ten voices, paired by voice, sentence and iteration.}
\label{fig:style-tag-bounds}
\end{figure}

\begin{figure}[!htbp]
\centering
\includegraphics[width=\linewidth]{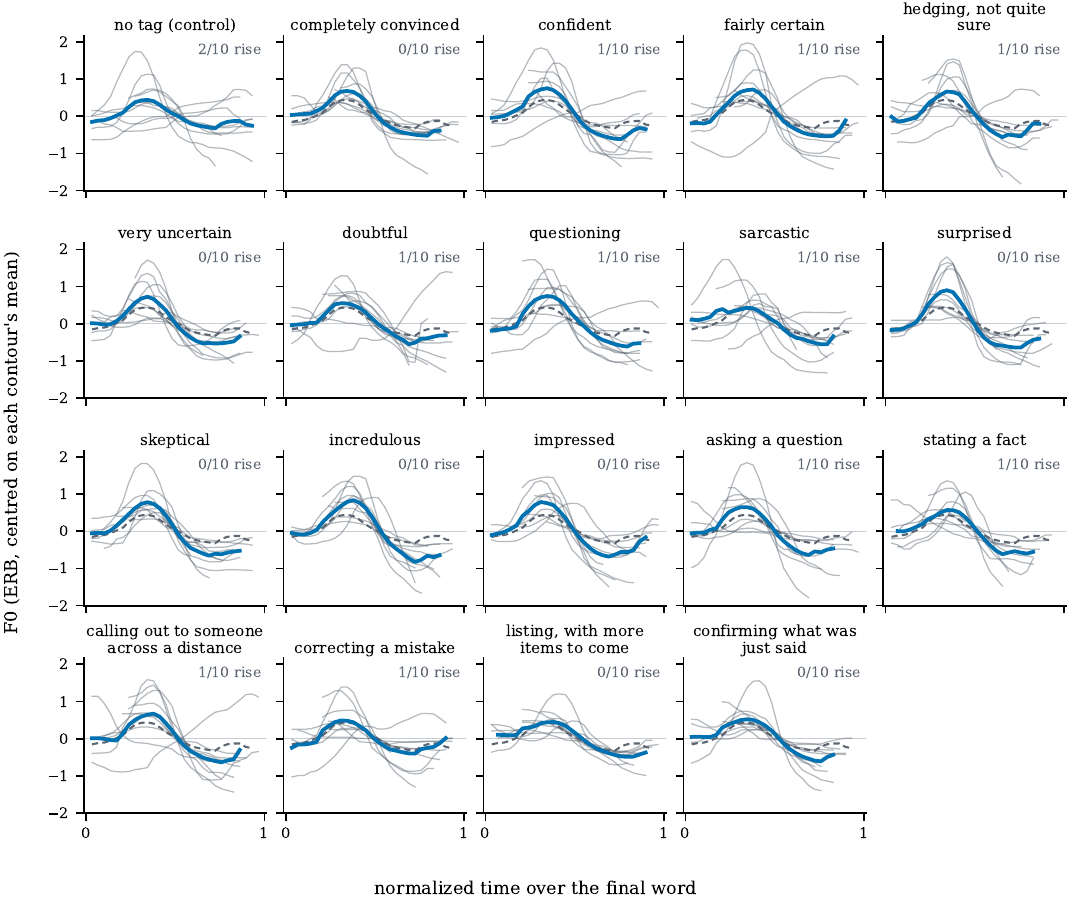}
\caption{Every style tag in the ten voices. One panel per tag, the untagged control first; each
panel draws one recording from each of the ten speakers, all prompted with that speaker's own
eight-tune voice prompt. Each contour is centred on its own mean, so that shape is compared
across voices whose registers differ. A speaker is shown on one carrier sentence, rotated across
speakers and the same in all of that speaker's panels, so that within a speaker only the tag
changes. Blue: the panel's mean where at least half of the recordings are voiced; dashed: the
control's mean, repeated in every panel; top right: how many of the ten rise past $+0.38$~ERB.}
\label{fig:style-tags}
\end{figure}

No tag moves the tune toward a rise (Figures~\ref{fig:style-tag-bounds} and~\ref{fig:style-tags}).
Each interval in Figure~\ref{fig:style-tag-bounds} is a confidence interval for one tag's mean
change, so the null hypothesis it tests is that the tag changes nothing: an interval that contains
zero is consistent with no effect, and one that stops short of the question mark's effect rules
out an effect of that size. Every tag's mean change in the excursion lies between $-0.38$ and
$0.00$~ERB. No Bonferroni interval excludes zero, in the excursion or in the word's highest or
lowest pitch; without the correction, $4$ of the $54$ intervals do, about what chance predicts
($2.7$): three lower the word
(\class{[impressed]} in its excursion and lowest pitch, \class{[confirming what was just said]} in
its lowest pitch), and \class{[questioning]} raises its highest pitch by $0.25$~ERB without
changing its excursion. Every interval also ends well short of the question mark's
effect of $2.09$ $[1.84, 2.34]$~ERB in the same voices: the widest $95\%$ interval of the
excursion reaches $\pm 0.80$~ERB, and the widest Bonferroni interval $\pm 1.11$. With voice-by-sentence blocks
the tag explains nothing ($F_{18,522} = 0.79$, $p = 0.71$); $48$ of $540$ tagged recordings rise,
against $4$ of $30$ untagged ones. The shape of the final-word contour is the same under every
tag: a rise-fall on the accented syllable followed by a fall to the end of the word.

\paragraph{Without the full stop.} A full stop is the standard end of an English sentence whatever
the speaker's credence, attitude or speech act, so the tags were tested with it; but it might also
override a tag. The same $570$ cells were therefore generated with no final punctuation. The
result is the same: every tag's mean change in the excursion lies between $-0.37$ and $+0.09$~ERB,
no Bonferroni interval excludes zero, in the excursion or in the word's highest or lowest pitch,
and the tag explains nothing ($F_{18,522} = 0.44$, $p = 0.98$); $102$ of $540$ tagged utterances
rise, against $4$ of $30$ untagged ones. CSM has no inline tags, so this control exists in S2 only.

\FloatBarrier
\subsection{Demonstration and lexical context}

Both controls were first run in CSM, in a single reference voice, and measured by the excursion
over an asserted final-word interval (the final two fifths of the voiced extent, which
\S\ref{app:robustness} shows recovers the aligned separation), with punctuation held fixed. They
were then repeated in S2, in the ten voices, on aligned final words.

\paragraph{Demonstration transfers, partly.} Three human recordings carrying one tune are given as
the voice prompt, and the conditions are a tune that falls (\class{hll}) against one that rises
(\class{lhh}). These two differ in all three tone symbols and are $3.07$~ERB apart in human
productions, more than the punctuation effect. In CSM the generations differ by $0.897$~ERB
$[0.646, 1.140]$, $45$ of $72$ rising under the rising demonstration: about a third of the
punctuation contrast, whose interval it excludes. In S2, with each speaker's own recordings of the
tune (none of the sentence being said) and paired by voice, sentence and seed, they differ by
$1.19$~ERB $[0.92, 1.46]$, $38$ of $90$ rising under the rising demonstration against $1$ of $90$
under the falling one: a little over half of the question mark's $2.09$~ERB in the same voices. A
demonstrated tune is not ignored, but it transfers only in part.

\paragraph{A lexical lead-in does nothing.} Four lead-ins presenting what follows as already
established (\textit{``As you know,''}, \textit{``Obviously,''}, \textit{``Of course,''},
\textit{``Everyone knows,''}) and four presenting it as news (\textit{``You won't believe this:''},
\textit{``Guess what:''}, \textit{``Listen to this:''}, \textit{``Here's the thing:''}) were
generated in CSM's prompt voice by the model itself, so that only the words differ; $576$
generations followed, $12$ per fragment per carrier sentence. The group contrast is $0.102$~ERB
$[-0.046, 0.247]$. In S2 the lead-in was written before the sentence in the same utterance, in the
ten voices: after a news lead-in the excursion is $0.10$~ERB \emph{lower} than after a known one
$[-0.24, 0.04]$, and $7$ of $120$ and $3$ of $120$ utterances rise, against $5$ of $30$ with no
lead-in.

\FloatBarrier
\section{Locating the tune}
\label{app:locating}

\FloatBarrier
\subsection{Where the unsteered conditions differ}
\label{app:where-differ}\label{app:prenuclear}\label{app:exclusions}

\begin{figure}[!htbp]
\centering
\begin{minipage}[c]{0.56\linewidth}
\centering
\includegraphics[width=\linewidth]{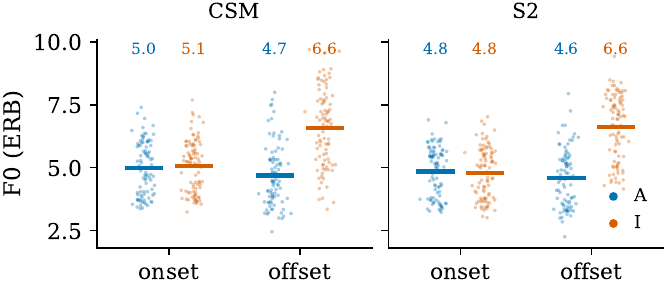}
\end{minipage}\hfill
\begin{minipage}[c]{0.4\linewidth}
\centering\small
\begin{tabular}{@{}l rr rr@{}}
\toprule
 & \multicolumn{2}{c}{CSM} & \multicolumn{2}{c}{S2} \\
\cmidrule(lr){2-3}\cmidrule(l){4-5}
region & I$-$A & $d$ & I$-$A & $d$ \\
\midrule
onset & $0.08$ & $0.08$ & $-0.06$ & $-0.05$ \\
offset & $1.89$ & $1.37$ & $2.04$ & $1.66$ \\
\bottomrule
\end{tabular}

\end{minipage}
\caption{Where the unsteered conditions differ on the final word, ten-speaker sets ($90$
recordings per condition and system). \emph{Left:} $F_0$ of the first and last fifth of the
word's voiced contour, one point per recording, with each condition's mean drawn and printed.
\emph{Right:} the separation of the condition means in ERB and as Cohen's $d$.}
\label{fig:final-word}
\end{figure}

A pair's two contours are compared, segment by segment, by their root-mean-square distance in ERB
at matched normalized time, dropping grid points where either is unvoiced. The same distance
between two generations of one condition (same voice and sentence, different iteration) is the
reference. Before the final word the two conditions are no further apart than two generations of
the same condition: the excess distance is $+0.09$~ERB $[-0.04, +0.23]$ in CSM and $+0.03$
$[-0.06, +0.12]$ in S2. Over the final word it is $+0.47$ $[+0.29, +0.66]$ and $+0.65$
$[+0.51, +0.79]$. Human imitations of \class{hll} and \class{lhh} by one speaker behave the same way
($+0.03$ before, $+0.92$ over). An intervention restricted to the final word is therefore not
missing a systematic difference elsewhere, and the cluster read-out is not separating the
conditions by register. Within the final word the difference is at the end
(Figure~\ref{fig:final-word}).

\FloatBarrier
\subsection{The swap}
\label{app:swap}\label{app:formal}

\begin{figure}[!htbp]
\centering
\includegraphics[width=\linewidth]{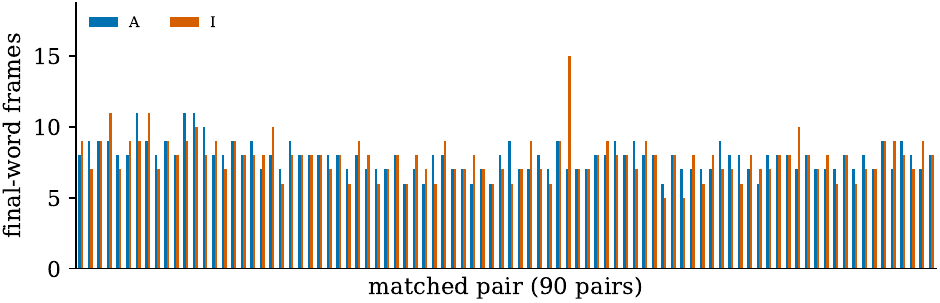}
\caption{The final-word length of the two recordings of each matched pair in frames, over the
$90$ CSM pairs of the ten-speaker set.}
\label{fig:alignment}

\vspace{10pt}
\includegraphics[width=0.62\linewidth]{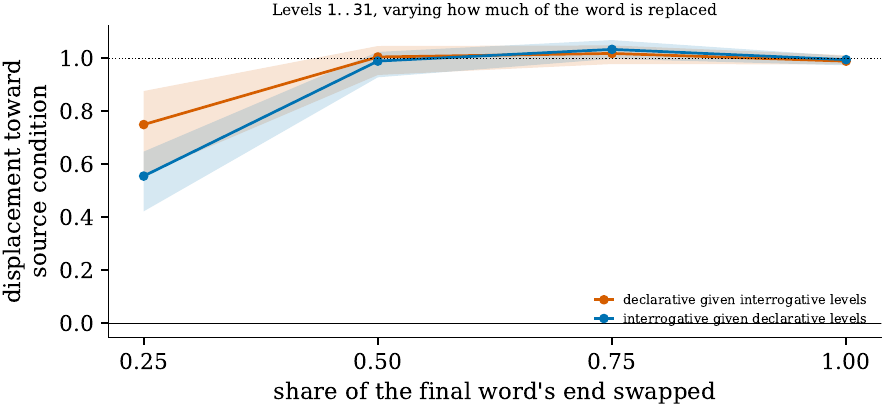}
\caption{Levels $1$--$31$ swapped over a varying share of the window, trimmed from the start so that
the final word's last frames are always included. Bands span the three read-outs.}
\label{fig:window-dose}

\vspace{10pt}
\includegraphics[width=0.62\linewidth]{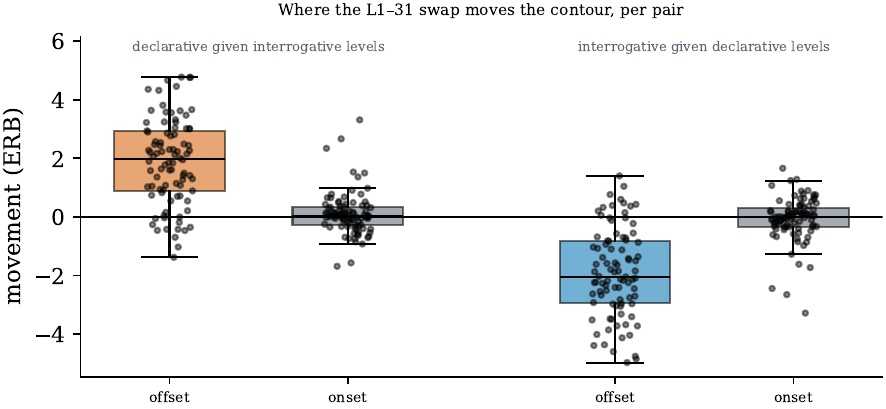}
\caption{Per-pair movement at the two ends of the final word under the $1$--$31$ swap. The offset
moves in opposite directions for the two swap directions; the onset does not move.}
\label{fig:positional-swap}
\end{figure}

The recording modified by the intervention is the \emph{target} (superscript $t$) and the recording
supplying the tokens is the \emph{source} ($s$). With $[\alpha, \beta]$ the final-word interval
from forced alignment and $f$ the frame rate, the final-word frames are
$N = \{\lfloor \alpha f \rfloor, \dots, \lceil \beta f \rceil\}$. Put $w = \min(|N_s|, |N_t|)$ and
take the last $w$ frames of each final word (the \emph{suffix} window, used for every result), or
the first $w$ (\emph{prefix}):
\begin{equation}
\begin{aligned}
  \text{\emph{suffix}}:\;\;
  &W_s = [\,\beta_s - w,\ \beta_s), &&W_t = [\,\beta_t - w,\ \beta_t); \\
  \text{\emph{prefix}}:\;\;
  &W_s = [\,\alpha_s,\ \alpha_s + w), &&W_t = [\,\alpha_t,\ \alpha_t + w).
\end{aligned}
  \label{eq:anchors}
\end{equation}
Let $q^t, q^s \in \{1,\dots,V\}^{T \times K}$ be the two token stacks, $L \subseteq \{0, \dots,
K-1\}$ a set of levels, and $\pi : W_t \to W_s$ the order-preserving bijection between the
windows. Then
\begin{equation}
  \bigl(\swapop_{L,W}(q^t;\, q^s)\bigr)_{t,k}
  \;=\;
  \begin{cases}
    q^s_{\pi(t),\,k} & \text{if } k \in L \text{ and } t \in W_t,\\[2pt]
    q^t_{t,k} & \text{otherwise,}
  \end{cases}
  \label{eq:swap}
\end{equation}
and the intervened audio is $\dec\bigl(\sum_k e^{(\swapop_{L,W}(q^t;\,q^s))_{t,k}}_k\bigr)$. Two
identities serve as implementation tests: $\swapop_{\emptyset,W}(q^t; q^s) = q^t$, and
$\bigl(\swapop_{\{0,\dots,K-1\},W}(q^t; q^s)\bigr)\big|_{W_t} = q^s\big|_{W_s}$. Matching equal
numbers of frames is the simplest policy that makes the swap well defined; interpolating codewords
instead would mean re-quantizing, which introduces a second approximation, or leaving token space.
In CSM's ten-speaker set final words are $8$ frames at the median and the two words of a pair
differ by $0.9$ frames on average (Figure~\ref{fig:alignment}), so the suffix window swaps $7.3$
frames and leaves little out: over the $90$ pairs, the suffix and the prefix windows each leave
$85$ final-word frames unswapped, less than one per pair. The swapped blocks differ in $98.2$ to
$99.7$ per cent of their token positions, except level $0$, which differs in $80.0$ per cent.

Where on the word the swap acts: exchanging levels $1$--$31$ over a shrinking share of the window,
always keeping the word's final frames, the final quarter of the window --- about two frames ---
already carries most of the effect (Figure~\ref{fig:window-dose}), and the full swap moves the
word's offset while leaving its onset in place (Figure~\ref{fig:positional-swap}).

\FloatBarrier
\subsection{Truncation}
\label{app:depth}

\begin{figure}[!htbp]
\centering
\includegraphics[width=0.9\linewidth]{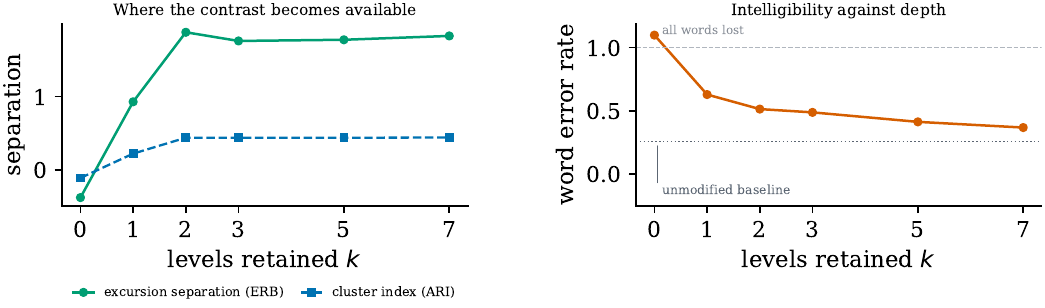}
\caption{Reconstruction from levels $0$--$k$, CSM. \emph{Left:} the conditions' separation by the
excursion and by an unsupervised cluster index against the design labels. At $k=0$ the
reconstruction carries no voiced material, so neither measure exists. \emph{Right:} word error
rate against depth, with the unmodified recordings' rate marked.}
\label{fig:depth}
\end{figure}

Reconstructing each recording from levels $0$ to $k$ \eqref{eq:partial}, the conditions' separation
appears at $k = 1$ and reaches the full stack's at $k = 2$, by the excursion and independently by
an unsupervised cluster index (Figure~\ref{fig:depth}). Level $0$ alone decodes to unvoiced,
unintelligible audio, and intelligibility returns over the same depths as the tune, so truncation
cannot separate tune from words; the swaps and the depth-model re-runs can.

\FloatBarrier
\subsection{What level 0 carries: the cross-sentence control}
\label{app:level0}

\begin{figure}[!htbp]
\centering
\includegraphics[width=\linewidth]{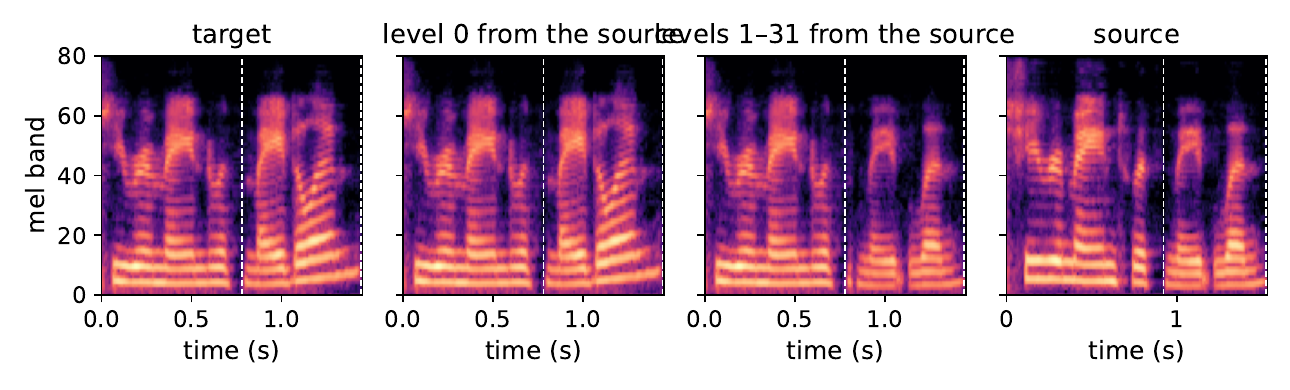}
\caption{One matched pair of the CSM ten-speaker set: the target, the target with level $0$ taken
from the source, the target with levels $1$--$31$ taken from the source, and the source. Log-mel
spectrograms; dashed lines bound the final word.}
\label{fig:level0-spectrogram}
\end{figure}

\paragraph{How much a level-0 swap changes the audio.}\label{app:level0-audio} In the matched
pairs of the ten-speaker set, a level-$0$ swap exchanges $80\%$ of the final word's level-$0$
tokens, yet the final word's excursion changes by $0.003$~ERB per pair on average (at most
$0.20$). Over the final word, resampled to a common length, the log-mel spectrogram of the swapped
recording differs from its target's by $1.7$~dB on average ($1.1$ to $3.1$ across pairs), against
$11.3$~dB for the swap of levels $1$--$31$ and $11.1$~dB between the target and the source
themselves (Figure~\ref{fig:level0-spectrogram}). Mimi's decoder, whose acoustic levels quantize
the whole frame in parallel with level $0$ (\S\ref{app:rvq}), thus writes the waveform almost
entirely from levels $1$--$31$: at the codec, a level-$0$ swap cannot move the tune because it
hardly moves anything. What level $0$ carries into the generation is the question of the depth
re-runs, and of the control below.

\paragraph{The cross-sentence control.}
Within a matched pair a level-$0$ swap moves tokens between two renditions of the same words, so an
unchanged transcript is expected. The control swaps between the assertions of two different
sentences by the same voice and iteration, over the final word, in both directions: $10$ voices
$\times$ $3$ sentence pairs $\times$ $3$ iterations $\times$ $2$ directions $= 180$ recordings per
variant, transcribed by Whisper-small and scored by whether the transcript names the target's final
word or the source's (Table~\ref{tab:level0}). At the codec, the source's level $0$ does not bring the
source's word with it, while the source's levels $1$--$31$ do: in the decoded waveform the word's
segmental content, like its tune, is carried by the levels the depth model emits. Inside the
depth model, the source's $q^0_t$ with the target's own $\vh_t$ garbles the word, and so does the source's
$\vh_t$ with the target's own $q^0_t$; only both together produce the source's word. The depth model
therefore writes the word from $q^0_t$ and $\vh_t$ jointly, and the tune from $\vh_t$ alone
\eqref{eq:route}.

\begin{table}[!htbp]
\centering\small
\begin{tabular}{@{}l r ccc@{}}
\toprule
CSM, final word of two sentences & $n$ & WER & target's name & source's name \\
\midrule
unmodified & $180$ & $0.26$ & $0.93$ & --- \\
swap level $0$ (codec) & $180$ & $0.20$ & $0.97$ & $0.00$ \\
swap levels $1$--$31$ (codec) & $180$ & $0.50$ & $0.00$ & $0.77$ \\
depth model: own $\vh_t$, source's $q^0_t$ & $180$ & $0.55$ & $0.06$ & $0.07$ \\
depth model: source's $\vh_t$, own $q^0_t$ & $180$ & $0.47$ & $0.13$ & $0.02$ \\
depth model: source's $\vh_t$ and $q^0_t$ & $180$ & $0.50$ & $0.02$ & $0.76$ \\
\bottomrule
\end{tabular}

\caption{The cross-sentence control, CSM ten-speaker set. \emph{Target's name}, \emph{source's
name}: share of transcripts containing the final word of the recording modified or of the one
supplying the tokens (the recognizer's spellings, e.g.\ \textit{Madeleine}, counted).}
\label{tab:level0}
\end{table}

\FloatBarrier
\subsection{Re-running the depth model offline}
\label{app:rerun}\label{app:backbone}

\paragraph{Why it can be done.} The generator samples a frame by computing $\vh_t$, drawing $q^0_t$
from the L0-head applied to it, and running the depth model over the codebook axis, starting from
$(\vh_t, e_0(q^0_t))$, to emit levels $1$ to $K-1$. The backbone attends over the whole prompt and
all audio emitted up to $t$; the depth model does not. In CSM its key-value cache is allocated with
as many positions as there are codebooks, its causal mask is $32 \times 32$, and its caches are
reset before every frame; in S2 the cache has ten positions, the mask is indexed by the
within-frame position, and every position a frame reads has been overwritten by that frame first.
Levels $1$ to $K-1$ of a frame are therefore a function of that frame's $(\vh_t, q^0_t)$ exactly.
The archives store $\vh_t$, the sampled stack and, in CSM, the level-$0$ logits for every frame, so
no backbone pass is needed.

\paragraph{How the re-run was verified.} In CSM, recomputing the L0-head on the stored $\vh_t$
reproduces the stored logits (largest disagreement $0.125$ on logits running to tens, argmax never
changing); with the depth loop made greedy, the offline re-run on the stored $(\vh_t, q^0_t)$
reproduces every level of every frame of three generations; and every frame recomputed in isolation
and in reverse order reproduces the forward pass token for token. In S2, the sampler restated with
an explicit random generator gives the library's probabilities on $2{,}000$ random logit vectors;
every stored token of levels $1$--$9$ lies in the set the re-run could have drawn from (with every
state shifted one frame back the rate falls to about one half, so the check can fail); and a
greedy re-run in order and in reverse agree token for token. A re-run at a fresh seed changes $95$
to $99$ per cent of the level-$1$-and-above tokens, which is why the re-run on the recording's own
inputs is reported as a null: token churn of that size has to be shown not to move the read-outs.

\paragraph{Decodability of the condition.} Table~\ref{tab:backbone} gives the linear decodability
of the condition from each level's codewords and from $\vh_t$ in CSM, on the ten-speaker set. The
backbone state decodes it over the final word and weakly before it, and is near chance at the
word's onset; the codewords of single levels, pooled over the final word, are near chance for
every level but level $2$. On an earlier set of $156$ CSM generations prompted with one recorded
reference voice instead of the ten speakers, level $1$ decoded the condition at $0.84$, so a
linear read-out of codewords does not transfer across voices, which is one reason the
localization rests on interventions rather than on decodability. In S2, on the ninety pairs, $\vh_t$
is at chance before and at the onset of the final word ($0.52$ and $0.55$) and decodes the
condition at $0.93$ over it.

\begin{table}[!htbp]
\centering\small
\begin{tabular}{lrrrrrr}
\toprule
& \multicolumn{3}{c}{pooled over the final word} & \multicolumn{3}{c}{backbone state $\vh_t$, by region} \\
\cmidrule(lr){2-4}\cmidrule(lr){5-7}
& accuracy & AUC & & region & accuracy & AUC \\
\midrule
level $0$ codewords & $0.522$ & $0.559$ & & before the final word & $0.661$ & $0.760$ \\
level $1$           & $0.528$ & $0.571$ & & final-word onset & $0.556$ & $0.690$ \\
level $2$           & $0.650$ & $0.703$ & & final word & $0.861$ & $0.927$ \\
level $3$           & $0.589$ & $0.603$ & & final-word offset & $0.839$ & $0.904$ \\
level $7$           & $0.467$ & $0.451$ & & final frame & $0.844$ & $0.929$ \\
$\vh_t$               & $0.861$ & $0.927$ & & whole recording & $0.878$ & $0.967$ \\
\bottomrule
\end{tabular}
\caption{Linear decodability of the condition, CSM ten-speaker set ($180$ recordings),
leave-one-carrier-sentence-out. The permutation null has mean accuracy $0.47$--$0.52$ and maximum
$0.54$--$0.64$ across cells, so values below about $0.64$ are not distinguishable from chance.}
\label{tab:backbone}
\end{table}

\FloatBarrier
\subsection{The second system}
\label{app:s2}

Fish Audio S2 \citep{fishaudio2026s2} pairs a $4$B autoregressive backbone with a $400$M depth
decoder over a $10$-level RVQ codec at ${\approx}21$~Hz; the sizes, and the token layout that
follows, are those of its code release. The backbone samples a token in its text
vocabulary from which level $0$ is obtained by a fixed offset, and the depth decoder samples levels
$1$--$9$. The S2 column of Table~\ref{tab:interventions} is the ninety matched pairs of the
ten-speaker set. Levels $1$--$9$ swap the tune ($\Delta_\Phi = 0.91$); level $1$ alone already
carries $0.71$ of it, against a fraction of that in CSM, where the residual stack has $31$ levels
instead of $9$ (\S\ref{app:rvq}); and level $0$ carries a small part ($0.09$ as a swap, $0.11$ as a
depth-model input). Swapping level $0$ in S2 also raises the word error rate from $0.130$ to
$0.228$ (Table~\ref{tab:interventions}). A plausible reason is that, unlike Mimi's, S2's
acoustic levels quantize the residual of level $0$, so a level-$0$ code taken from another
recording no longer matches the residual the other levels encode.
Depth $0$ has no counterpart: S2's codec decodes level $0$ and the residual by separate paths and
does not accept a one-level stack. The steering directions are refitted on S2's own states, leaving
one carrier sentence out; the probe and mean-difference directions agree to a cosine of $0.50$ to
$0.55$ across folds (\S\ref{app:offline}).

\paragraph{Frame rate.} S2's final-word windows were computed at $21.0$ frames per second instead of
its codec's $21.53$ ($44.1$~kHz audio, $2048$ samples per frame). Over the $180$ recordings of the
ten-speaker set, the window, $14$ frames at the median, therefore ends one frame early in $73\%$
of recordings and starts one frame early in $39\%$; it is never off by more than one frame and
always covers at least $92\%$ of the word ($95\%$ on average). The shift applies equally to every
S2 intervention, to the steering onsets and to the human-derived vectors, and may account for part
of the full swap's shortfall from $1$ ($\Delta_\Phi = 0.91$).

\FloatBarrier
\subsection{Intelligibility}
\label{app:wer}

Word error rate is Whisper-small's against the prompt, both strings lower-cased and stripped of
punctuation, since the recognizer punctuates by what it hears and would otherwise count a
punctuation mark as a substituted word, differently by condition. The generators' own output is not
perfectly intelligible --- $0.26$ in CSM and $0.13$ in S2 on the ten-speaker sets, largely misheard
names (\textit{Madeleine} for \textit{Madelyn}): each sentence is three or four words of which
one is a proper name, so a single misheard name is a quarter of the utterance --- and every rate is
read against the unmodified recordings' rate. In CSM the depth-model re-run on the target's own
inputs raises the rate from $0.26$ to $0.34$, so the re-run blocks of Table~\ref{tab:interventions}
are read against that row, not against the unmodified one.

\FloatBarrier
\section{Offline steering}
\label{app:offline}\label{app:vq}

\begin{figure}[!htbp]
\centering
\includegraphics[width=\linewidth]{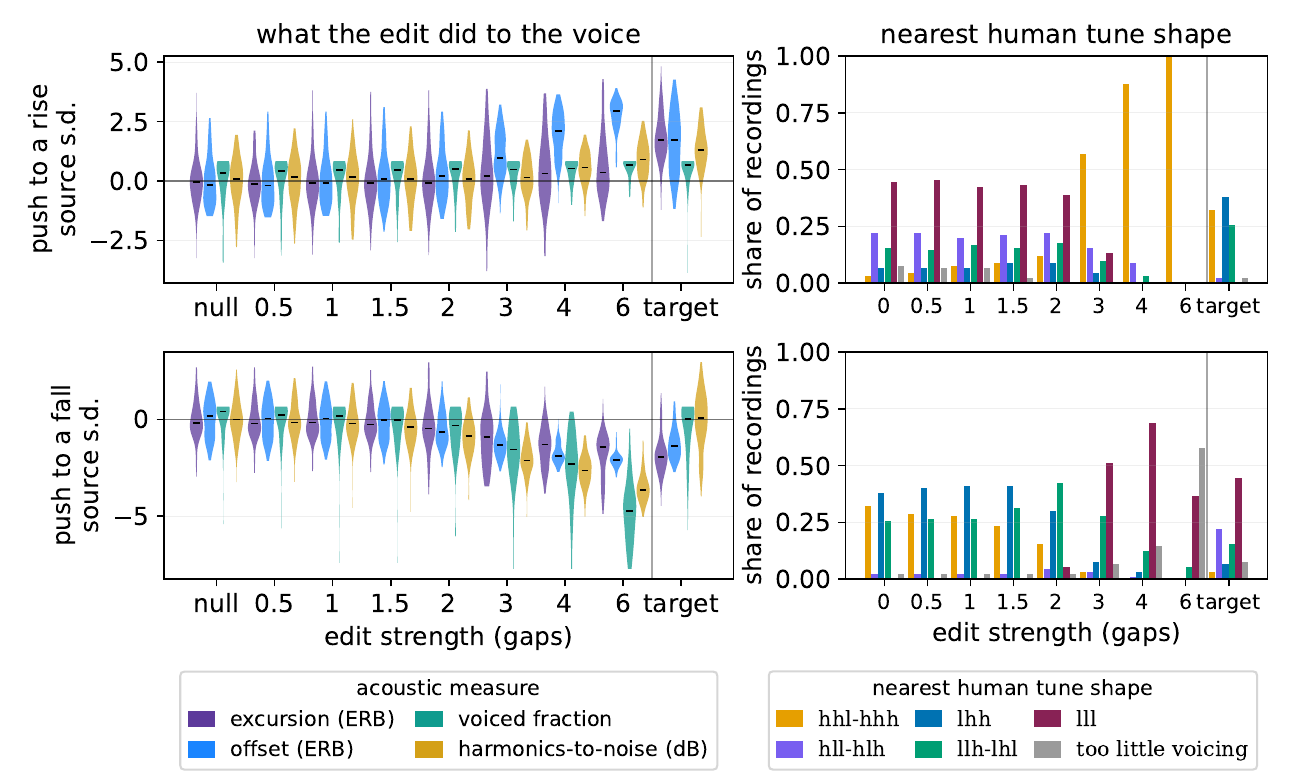}
\caption{Steering along the punctuation vector $\hvw$ offline, CSM, ten-speaker set. One row
per direction of the push: assertions pushed toward a rise (top) and questions toward a fall
(bottom), $90$ recordings per strength. \emph{Left:} what the edit does to the voice. Each violin
is one acoustic measure over the final word --- excursion, offset, voiced fraction, harmonics-to-noise
ratio --- in units of the source condition's standard deviation, centred on the source's own
re-run without an edit (\emph{null}); the rightmost group is the unmodified target condition. A
strength is therefore read as a position between the source and the target. \emph{Right:} the
nearest of the five human tune shapes for the same recordings (\S\ref{sec:measures}), with the
target condition's own distribution at the right; grey, words with too little voicing for a
contour (\S\ref{app:shapes}). The strength axis is in gaps \eqref{eq:gap}. The edit is
made before the normalization, $\mathrm{norm}(\vx_t + \alpha g \hvw)$, the form used in the rest
of the paper; the S2 version is Figure~\ref{fig:offline-s2}.}
\label{fig:offline}
\end{figure}

\paragraph{Directions.} Offline, every edit is made before the backbone's normalization,
$\mathrm{norm}(\vx_t + \alpha g \hvw)$, with $\hvw$ fitted on the unnormalized state $\vx_t$, which
keeps every edited state on the surface the heads were trained on. The mean-difference direction
is the difference of means of $\vx_t$ over every final-word frame of each condition
\eqref{eq:gap}; for Table~\ref{tab:interventions} it is refitted leaving one carrier sentence out,
and for Figure~\ref{fig:offline} it is fitted on all ten voices. The discriminative direction of
Table~\ref{tab:interventions} is a logistic regression on $\vx_t$ pooled over the final word, mapped
out of the standardized space and normalized. In S2 both are fitted with the two coordinates that
hold $97\%$ of $\lVert \vx_t\rVert^2$ set to zero. Held out, the probe classifies the condition at
$0.83$ to $0.88$ in CSM and $0.90$ to $0.97$ in S2, the difference of means at $0.82$ to $0.88$ and
$0.82$ to $0.87$; the two directions agree to a cosine of $0.64$ to $0.68$ in CSM and $0.50$ to
$0.55$ in S2. One gap is $10.7$ to $11.4$ state units in CSM and $152$ to $181$ in S2.

To see why the probe direction fails, its component orthogonal to $\hvw$ was steered on its own in
CSM, at the length it has inside the probe edits of $2$ and $4$ gaps ($16.5$ and $33$ state units):
it moves the tune by $\Delta_\Phi = 0.01$ and $0.02$ (mean over read-outs and directions), while the
probe edit of $4$ gaps, which adds $2.6$ gaps along $\hvw$ to that component, moves it by $0.05$,
and the difference of means alone at $3$ gaps by $0.61$. The probe's edit is therefore not simply a
weaker version of the mean difference: its component along $\hvw$ would on its own move the tune
about half of the way, its orthogonal component does nothing, and together they cancel. The
contrast between the two directions resembles the one \citet{marks2024geometry} report for
directions of truth in language models; in their case, as in \citet{li2023iti}, difference-of-means
is also better at steering. We leave the mechanism open.

\paragraph{The cost of the edit.} Four forms of the offline edit were run on the ten-speaker set:
after the normalization; the same with $q^0_t$ redrawn from the edited state before the depth model
runs; before the normalization; and, in S2, before the normalization with the two coordinates that
hold $97\%$ of $\lVert \vx_t\rVert^2$ excluded. At the strength that matches the target condition's
mean offset, all four cost about the same (Table~\ref{tab:vq}). Re-running the depth model on
unedited states costs about $2$~dB of harmonicity by itself, so most of the loss is from the edit.
What the four share is that the word's level-$0$ content, its duration and the frames after it are
held at the value of the question, which is what the in-loop protocol of \S\ref{sec:human} releases.

\begin{table}[!htbp]
\centering\small
\setlength{\tabcolsep}{4pt}
\begin{tabular}{@{}ll rrrr rrrr@{}}
\toprule
& & \multicolumn{4}{c}{\class{I}$\to$\class{A}} & \multicolumn{4}{c}{\class{A}$\to$\class{I}} \\
\cmidrule(lr){3-6}\cmidrule(l){7-10}
& Edit & $\alpha$ & offset & voiced & HNR & $\alpha$ & offset & voiced & HNR \\
\midrule
CSM & natural target & & $4.69$ & $0.86$ & $15.3$ & & $6.58$ & $0.93$ & $17.7$ \\
& depth re-run only & $0$ & $6.58$ & $0.92$ & $15.5$ & $0$ & $4.70$ & $0.85$ & $13.1$ \\
& after the norm & $8$ & $4.68$ & $0.74$ & $9.7$ & $10$ & $6.54$ & $0.91$ & $14.1$ \\
& after the norm, $q^0_t$ redrawn & $8$ & $4.65$ & $0.74$ & $9.9$ & $10$ & $6.75$ & $0.92$ & $14.3$ \\
& before the norm & $30$ & $4.65$ & $0.72$ & $9.4$ & $40$ & $7.04$ & $0.92$ & $15.1$ \\
S2 & natural target & & $4.59$ & $0.83$ & $16.7$ & & $6.63$ & $0.89$ & $17.8$ \\
& depth re-run only & $0$ & $6.60$ & $0.88$ & $16.1$ & $0$ & $4.60$ & $0.81$ & $14.6$ \\
& after the norm & $10$ & $4.53$ & $0.71$ & $10.2$ & $8$ & $6.89$ & $0.85$ & $15.0$ \\
& after the norm, $q^0_t$ redrawn & $7$ & $4.62$ & $0.69$ & $10.4$ & $7$ & $7.16$ & $0.87$ & $17.0$ \\
& before the norm & $800$ & $4.42$ & $0.71$ & $9.9$ & $600$ & $6.46$ & $0.82$ & $14.5$ \\
& before the norm, two coordinates out & $660$ & $4.56$ & $0.71$ & $10.2$ & $500$ & $6.48$ & $0.83$ & $14.6$ \\
\bottomrule
\end{tabular}
\caption{Voice quality of the offline edits at matched offset, ten-speaker set, $90$ targets per
direction. \emph{Offset}: mean offset level of the steered word (ERB); \emph{voiced}: voiced
fraction; HNR in dB. Each row is the grid value of $\alpha$ (state units) whose mean offset is
closest to the natural target; one gap is about $3.1$ state units after the normalization in CSM
and $2.3$ in S2, and before it $10.6$ in CSM and $166$ in S2 with the two coordinates out; \emph{natural target} gives that condition's unmodified values, and
\emph{depth re-run only} the steered condition's recordings re-run without an edit.}
\label{tab:vq}
\end{table}

\begin{figure}[!htbp]
\centering
\includegraphics[width=0.93\linewidth]{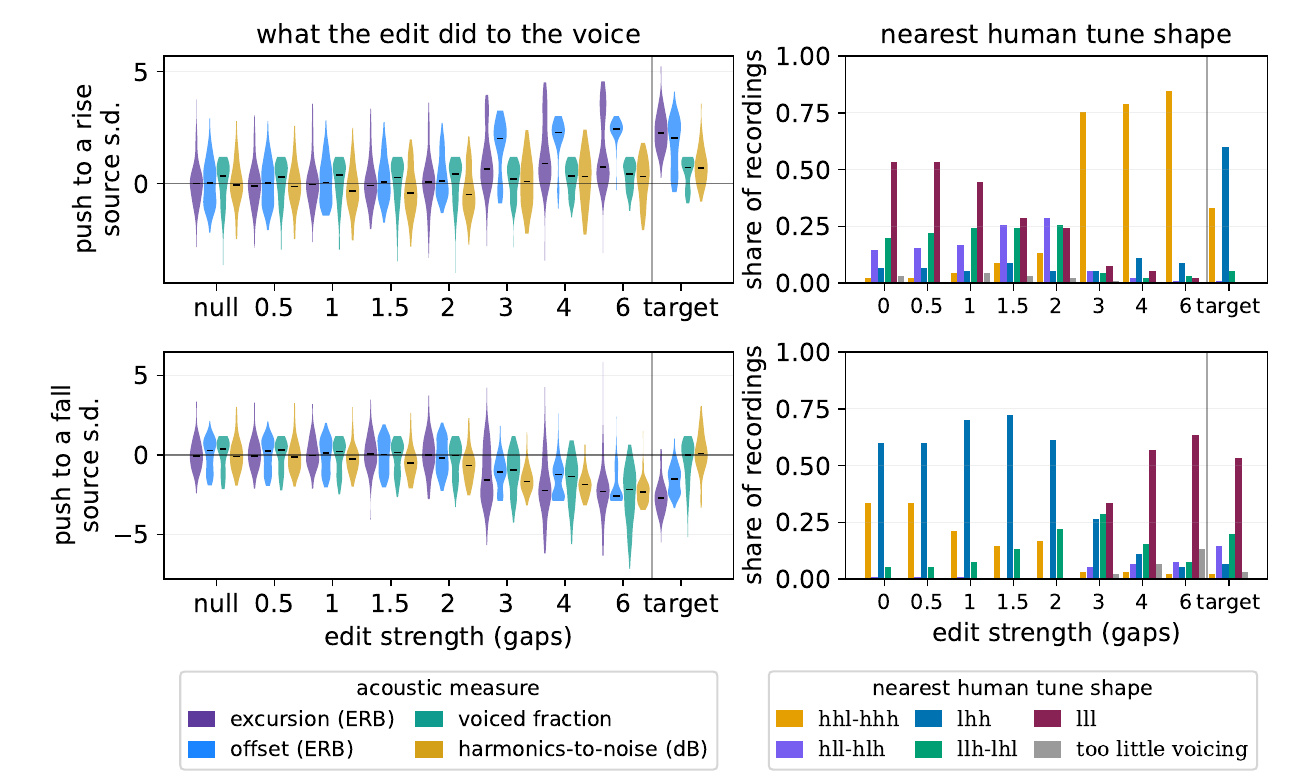}
\caption{Steering along the punctuation vector offline in S2, drawn as Figure~\ref{fig:offline}.}
\label{fig:offline-s2}
\end{figure}

\FloatBarrier
\section{Steering inside the generation loop}
\label{app:inloop}\label{app:directions}

\paragraph{The vectors.} Each recording is fed as a voice prompt would be --- its transcript from
the alignment, which carries no punctuation, then its own codec tokens --- and the pre-normalization
state $\vx_t$ is recorded at every frame. The state assigned to frame $i$ is the one computed after
reading frame $i-1$, the state the L0-head reads to choose frame $i$; the capture checks
this against the model's own step. Frames of the final word are pooled by tune group and the
difference of the two means is taken. In S2 the two coordinates with the largest root-mean-square,
which hold $95$ to $98\%$ of $\lVert x \rVert^2$ at every frame (the rogue or outlier dimensions
reported for transformer language models, \citealp{timkey2021rogue,dettmers2022int8}), are zeroed first, since an edit along them acts as
a gain change. The \textsf{tune} vector separates held-out human recordings at $0.81$ (CSM) and
$0.74$ (S2) accuracy; leave-one-speaker-out cosines with the pooled vector are $0.97$--$0.99$.
Frames are pooled, not recordings, so a long final word contributes more than a short one, and the
groups are not exactly balanced across speakers.

\paragraph{Alignment with the punctuation vector.} The \textsf{tune} vector's cosine with $\hvw$
is $0.66$ in CSM and $0.44$ in S2. As a null, the same difference of means was fitted $200$ times
on the same recordings with their tune labels shuffled within each speaker and sentence, so that
each half keeps the speakers, sentences and sizes of the real contrast: the cosine reaches at most
$0.31$ in CSM and $0.19$ in S2 ($95$th percentile of its absolute value; the shuffled directions'
mean is $0.18$ and $-0.04$). Among the human-fitted directions, the one fitted on the edge tones
alone is the most aligned ($0.78$, $0.80$), and in CSM the one fitted on the pitch accent alone the
least ($0.28$; $0.58$ in S2), as expected from where the two conditions differ (\S\ref{sec:where}).
The human-fitted directions are computed on states that follow the audio already read, so they may
carry the recent pitch movement as much as a planned tune; the alignment shows a shared
representation of intonation, not of tune categories.

\paragraph{Leave-one-speaker-out steering.} For each system and each speaker, the \textsf{tune}
vector was refitted on the other nine speakers' human recordings and used to steer that speaker's
$18$ targets in the loop, at the two strengths the paper reads, in units of the refitted vector's
own gap (Table~\ref{tab:loso}). Flip rates are within the in-sample vector's intervals in every
cell, and so is the voice quality of the steered words.

\begin{table}[!htbp]
\centering\small
\begin{tabular}{llcccc}
\toprule
system & push & vector & flip rate & voiced fraction & HNR (dB) \\
\midrule
CSM & rise, $2.5$ gaps & held out & $0.60$ $[0.50, 0.69]$ & $0.97$ & $17.7$ \\
    &                  & in-sample & $0.56$ $[0.48, 0.64]$ & $0.98$ & $18.1$ \\
    & fall, $1.5$ gaps & held out & $0.60$ $[0.53, 0.67]$ & $0.90$ & $14.0$ \\
    &                  & in-sample & $0.56$ $[0.46, 0.66]$ & $0.89$ & $14.3$ \\
S2  & rise, $2.5$ gaps & held out & $0.76$ $[0.67, 0.83]$ & $0.92$ & $19.6$ \\
    &                  & in-sample & $0.72$ $[0.62, 0.82]$ & $0.92$ & $19.0$ \\
    & fall, $1.5$ gaps & held out & $0.58$ $[0.43, 0.73]$ & $0.88$ & $14.5$ \\
    &                  & in-sample & $0.57$ $[0.38, 0.74]$ & $0.87$ & $14.2$ \\
\bottomrule
\end{tabular}
\caption{Leave-one-speaker-out steering: flip rates with $95\%$ intervals bootstrapped over
speakers, and median voice quality of the steered final words, $90$ targets per cell. \emph{Held
out}: each speaker steered with a vector fitted on the other nine; \emph{in-sample}: the vector
fitted on all ten.}
\label{tab:loso}
\end{table}

\paragraph{Replaying the prefix.} A target is regenerated with its archived seed and its own voice
prompt, the eight recordings the prompt plan assigned to that voice. Frames before the onset of the
final word are computed as in generation, with their random draws made, and their sampled tokens
are replaced by the archived ones, so that the backbone's cache and the random stream are those of
the original run when the edit begins. In CSM the frame step is a copy of the generator's, which a
test on a small instance of the model code shows to reproduce it draw for draw, and at $\alpha = 0$
every run reproduces its original recording frame for frame. In S2 every run reproduces itself, but
for the first prompt of each sentence the replay does not match the original recording, which was
the first generated after a change of sentence; the original driver seems to have carried state
from one generation to the next. S2's in-loop results are therefore read against the replayed
$\alpha = 0$ run rather than the original recordings.

\paragraph{Strength, window and measures.} The grid is $1.5$, $2.5$, $3.5$ and $4.5$ times each
vector's own gap, the distance between the mean states of its two groups of human recordings;
CSM's \textsf{tune} gap is $10.19$, so the grid is $\alpha \approx 15.3$ to $45.8$ in state units.
(Along that direction, the model's own unsteered assertions and questions are $7.10$ apart.) Steering from the onset to the end of the utterance, the model often fails to end
it; with the final-word window the steered utterances stay within about two frames of the
original's length. Steered recordings are measured on their own forced alignment. Against each
recording's own unsteered run, the mean change in word error rate is $+0.062$ and $-0.010$ in CSM
for the two directions and $+0.011$ and $-0.004$ in S2, with a median of exactly zero in all four.

\paragraph{Choosing the strength.}\label{app:strength} For each system and push, the steered
recordings at each strength are compared with the unmodified recordings of the target condition on
the four measures of Figures~\ref{fig:alpha} and~\ref{fig:alpha-s2}. On each measure the distance
is the $1$-Wasserstein distance between the two distributions, divided by the standard deviation of
the steered recordings' own unsteered runs, the scale the figures are drawn in; the criterion $D$
is the mean over the four measures, and the strength read is the one that minimizes it
(Table~\ref{tab:strength}). A strength that carries the pitch past the target, or that buys the
pitch movement with a loss of voicing or harmonicity the target does not show, is penalised by the
same criterion. The choice does not depend on the distance: the mean Kolmogorov--Smirnov statistic
and the mean absolute difference of means, in the same units, select the same four strengths. Nor
does it depend on the speakers it is measured on: chosen on nine voices and applied to the tenth,
for each voice in turn, it selects the same strength in all ten folds of all four cells, so the
held-out flip rates equal those reported.

\begin{table}[!htbp]
\centering\small
\begin{tabular}{ll cccc}
\toprule
system & push & $1.5$ & $2.5$ & $3.5$ & $4.5$ \\
\midrule
CSM & toward a rise & $0.48$ & $\mathbf{0.30}$ & $0.43$ & $0.76$ \\
CSM & toward a fall & $\mathbf{0.37}$ & $0.89$ & $1.50$ & $1.95$ \\
S2  & toward a rise & $0.57$ & $\mathbf{0.38}$ & $0.54$ & $0.70$ \\
S2  & toward a fall & $\mathbf{0.48}$ & $0.82$ & $1.14$ & $1.45$ \\
\bottomrule
\end{tabular}
\caption{The strength criterion $D$, the mean over excursion, offset, voiced fraction and HNR of
the Wasserstein distance between the steered recordings and the target condition's unmodified
recordings, in units of the unsteered runs' standard deviation, by strength in gaps; $90$
recordings per cell. Bold: the minimum, the strength read.}
\label{tab:strength}
\end{table}

\FloatBarrier
\subsection{Flip rates}
\label{app:flipci}

A \emph{flip} counts a recording whose steered excursion ends on the target side of $+0.38$~ERB when
its own unsteered run did not (Table~\ref{tab:flips}). Each cell is ten speakers $\times$ three
sentences $\times$ three iterations, so intervals resample \emph{speakers}, $2000$ draws; per-speaker
rates within one cell run from $0.22$ to $1.00$, and differences smaller than about $0.15$ are not
interpretable. For the \textsf{tune} vector the velocity and edge-probe read-outs agree with the
excursion to within $0.12$ at every strength, and each puts at most $0.11$ of recordings on the
wrong side. They diverge more for the other two vectors: pushed toward a fall from $2.5$ gaps, the
velocity counts up to $0.34$ fewer flips than the excursion, and pushed toward a rise, the
\textsf{accent} vector moves up to $0.23$ of CSM's recordings the wrong way by the velocity
read-out ($0.17$ by the excursion).

\begin{table}[!htbp]
\centering\footnotesize
\setlength{\tabcolsep}{3pt}
\begin{tabular}{llcccc}
\toprule
intervention & direction & $1.5$ & $2.5$ & $3.5$ & $4.5$ \\
\midrule
\multicolumn{6}{l}{\emph{CSM}} \\
\textsf{tune}   & \class{A}$\to$\class{I} & $0.36$ $[0.22, 0.49]$ & $0.56$ $[0.48, 0.64]$ & $0.51$ $[0.42, 0.60]$ & $0.64$ $[0.52, 0.76]$ \\
                & \class{I}$\to$\class{A} & $0.56$ $[0.45, 0.67]$ & $0.82$ $[0.77, 0.88]$ & $0.81$ $[0.71, 0.90]$ & $0.76$ $[0.69, 0.82]$ \\
\textsf{edge}   & \class{A}$\to$\class{I} & $0.30$ $[0.21, 0.39]$ & $0.56$ $[0.49, 0.62]$ & $0.53$ $[0.45, 0.61]$ & $0.36$ $[0.29, 0.43]$ \\
                & \class{I}$\to$\class{A} & $0.63$ $[0.53, 0.74]$ & $0.78$ $[0.74, 0.83]$ & $0.69$ $[0.59, 0.80]$ & $0.58$ $[0.51, 0.67]$ \\
\textsf{accent} & \class{A}$\to$\class{I} & $0.16$ $[0.09, 0.23]$ & $0.19$ $[0.09, 0.31]$ & $0.30$ $[0.23, 0.38]$ & $0.42$ $[0.35, 0.48]$ \\
                & \class{I}$\to$\class{A} & $0.08$ $[0.02, 0.15]$ & $0.17$ $[0.09, 0.26]$ & $0.33$ $[0.24, 0.43]$ & $0.34$ $[0.24, 0.42]$ \\
\midrule
\multicolumn{6}{l}{\emph{S2}} \\
\textsf{tune}   & \class{A}$\to$\class{I} & $0.51$ $[0.43, 0.59]$ & $0.72$ $[0.62, 0.83]$ & $0.79$ $[0.70, 0.88]$ & $0.71$ $[0.61, 0.82]$ \\
                & \class{I}$\to$\class{A} & $0.57$ $[0.38, 0.74]$ & $0.80$ $[0.70, 0.90]$ & $0.84$ $[0.79, 0.90]$ & $0.82$ $[0.68, 0.93]$ \\
\textsf{edge}   & \class{A}$\to$\class{I} & $0.51$ $[0.38, 0.63]$ & $0.71$ $[0.61, 0.80]$ & $0.81$ $[0.77, 0.86]$ & $0.82$ $[0.77, 0.88]$ \\
                & \class{I}$\to$\class{A} & $0.60$ $[0.44, 0.77]$ & $0.84$ $[0.80, 0.90]$ & $0.82$ $[0.73, 0.90]$ & $0.58$ $[0.40, 0.72]$ \\
\textsf{accent} & \class{A}$\to$\class{I} & $0.12$ $[0.03, 0.23]$ & $0.13$ $[0.08, 0.19]$ & $0.18$ $[0.10, 0.26]$ & $0.43$ $[0.32, 0.53]$ \\
                & \class{I}$\to$\class{A} & $0.12$ $[0.04, 0.23]$ & $0.33$ $[0.17, 0.53]$ & $0.53$ $[0.37, 0.71]$ & $0.74$ $[0.61, 0.87]$ \\
\bottomrule
\end{tabular}
\caption{Flip rates by intervention and strength in gaps, with $95\%$ intervals bootstrapped over
speakers; $90$ recordings per cell.}
\label{tab:flips}
\end{table}

\paragraph{The no-edit baseline.} Under deterministic kernels the $\alpha = 0$ replay reproduces the
source, so it says nothing about how often a \emph{fresh} draw from the model would cross the
boundary unaided. Regenerating the whole set four times per target with the seed offset by the
repeat, the same event occurs between draws in $0.203$ $[0.144, 0.280]$ of CSM's assertions and
$0.148$ $[0.104, 0.200]$ of S2's, and in $0.067$ and $0.030$ of their questions: the models'
assertions are flat, so a small perturbation moves them across. The \textsf{tune} and \textsf{edge}
rates exceed that baseline by $0.35$ to $0.57$ at the strengths the paper reads, in both systems and
directions.

\FloatBarrier
\subsection{Two targets, and which one the steering meets}
\label{app:targets}

\begin{figure}[!htbp]
\centering
\includegraphics[width=0.8\linewidth]{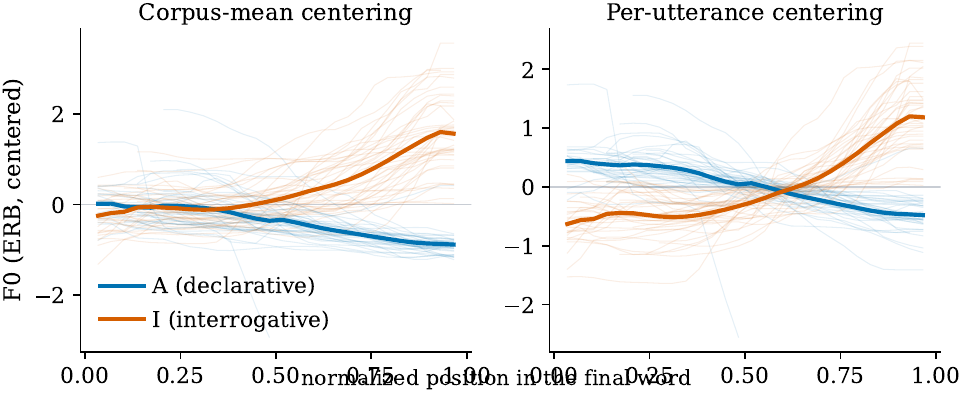}
\caption{The pilot: mean final-word contours of CSM's two conditions in a single voice ($84$
recordings), centred on the mean of all contours (left) and on each contour's own mean
(right), with individual recordings faint.}
\label{fig:contours}
\end{figure}

A steering vector fitted on one contrast and applied to another has two defensible targets. The
\emph{nominal} target is the tune the vector was fitted to produce: pushing along the \textsf{tune}
vector should produce \class{lhh}, and against it \class{hll}. The \emph{empirical} target is the
distribution of tunes the model itself produces in the condition being steered toward. We devised
the intervention in terms of the nominal target, and a pilot analysis of CSM's two conditions in a
single voice, whose mean contours fall and rise as those two tunes do, seemed to support it
(Figure~\ref{fig:contours}). The ten-speaker material shows more variability: neither endpoint
describes it (Table~\ref{tab:unsteered-classes}). The models' assertions are mostly \class{lll}, and
their questions divide between \class{lhh} and the high plateau \class{hhl}/\class{hhh}.

\begin{table}[!htbp]
\centering\small
\setlength{\tabcolsep}{4pt}
\begin{tabular}{llcccccc}
\toprule
system & unsteered condition & \class{hhl}/\class{hhh} & \class{hll}/\class{hlh} & \class{lhh} & \class{llh}/\class{lhl} & \class{lll} & too little voicing \\
\midrule
CSM & assertions (\class{A}) & $0.033$ & $0.233$ & $0.100$ & $0.122$ & $\mathbf{0.478}$ & $0.033$ \\
CSM & questions (\class{I})  & $0.300$ & $0.022$ & $\mathbf{0.400}$ & $0.256$ & $0.000$ & $0.022$ \\
S2  & assertions (\class{A}) & $0.011$ & $0.144$ & $0.033$ & $0.211$ & $\mathbf{0.589}$ & $0.011$ \\
S2  & questions (\class{I})  & $0.344$ & $0.011$ & $\mathbf{0.578}$ & $0.067$ & $0.000$ & $0.000$ \\
\bottomrule
\end{tabular}
\caption{The nearest human tune shape of the unsteered ten-speaker recordings, as shares of all
$90$ recordings per condition; the last column counts words with too little voicing for a contour
(\S\ref{app:shapes}).}
\label{tab:unsteered-classes}
\end{table}

\begin{figure}[!htbp]
\centering
\includegraphics[width=0.93\linewidth]{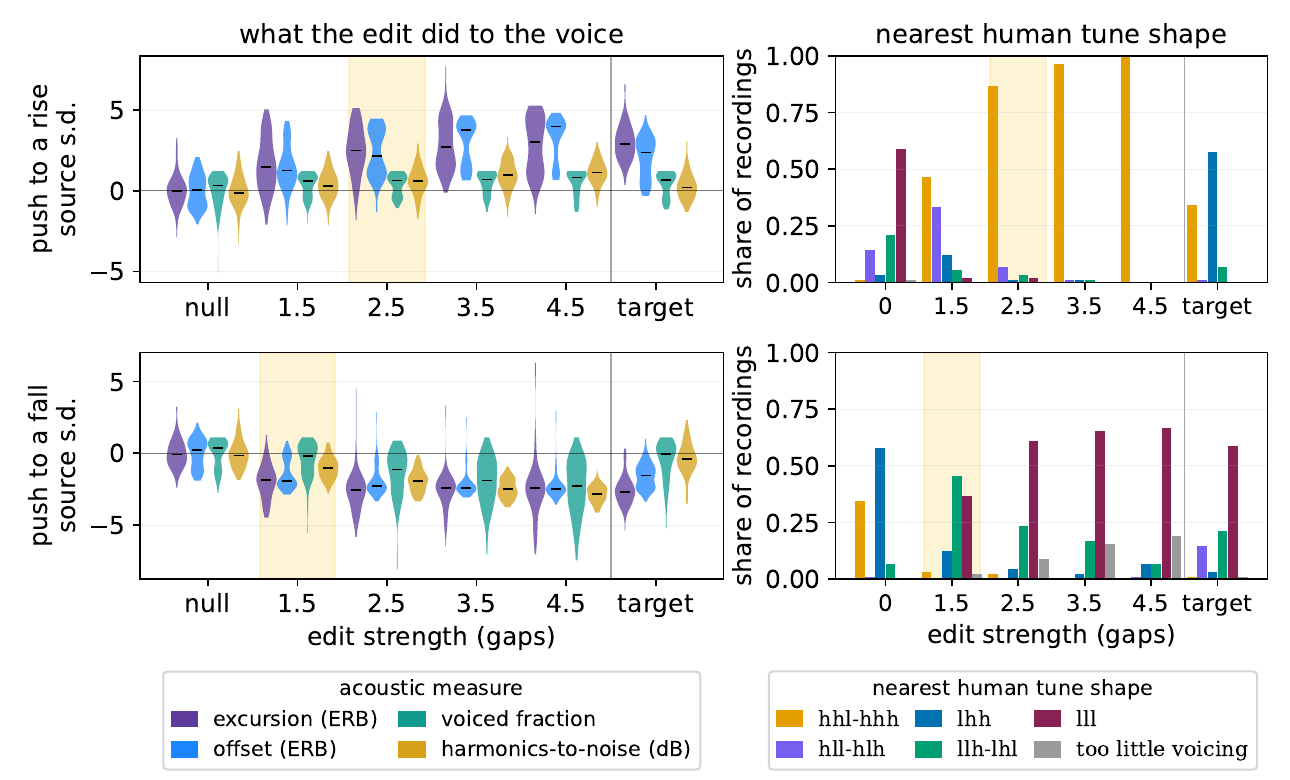}
\caption{Steering along the \textsf{tune} vector inside the loop in S2, drawn as
Figure~\ref{fig:alpha}. The push to a rise reaches the high plateau rather than \class{lhh}, the
most frequent shape of S2's own questions ($0.58$), which the steered set holds at most $0.12$ of,
and less with strength.}
\label{fig:alpha-s2}
\end{figure}

Against the nominal target the intervention fails: at any strength, at most $0.17$ of the
recordings pushed toward a rise are nearest \class{lhh}, and at most $0.02$ of those pushed
toward a fall are nearest \class{hll}/\class{hlh}. Against the empirical target it partly
succeeds, and differently in the two directions (Figures~\ref{fig:alpha} and~\ref{fig:alpha-s2}).
Pushed toward a fall, the recordings move to \class{lll}, the models' most frequent assertion
shape, and to the mixed-edge \class{lhl}/\class{llh}. Pushed toward a rise, they move to the
high plateau, increasingly so with strength; that shape holds about a third of the models' own
questions ($0.30$ in CSM, $0.34$ in S2), while their most frequent question shape, \class{lhh},
is rarely reached. The edit moves a recording toward where the target condition lives without
reconstituting the tune its vector was fitted on.

\FloatBarrier
\subsection{Quality over every steered recording}
\label{app:quality}

\begin{figure}[!htbp]
\centering
\includegraphics[width=\linewidth]{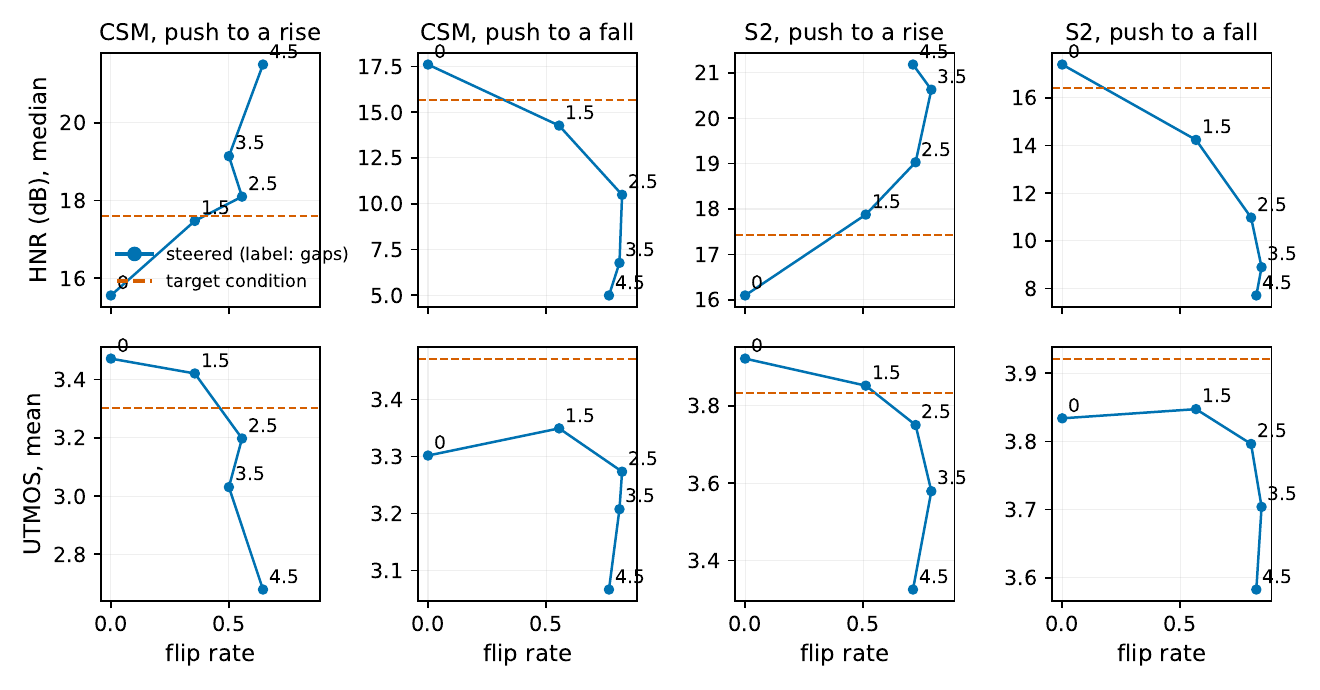}
\caption{Flip rate against quality, over every steered recording of each strength of the
\textsf{tune} vector in the loop (numbers: strength in gaps; $0$: the unsteered run). Top:
median harmonicity of the final word. Bottom: mean UTMOS of the utterance. Dashed: the target
condition's own recordings.}
\label{fig:flip-quality}
\end{figure}

Conditioning on a flip selects recordings that could be measured and that moved, so this first
reports quality over every steered recording of a strength, flipped or not, runaway generations
included wherever the final word could still be measured (Figure~\ref{fig:flip-quality}). Two
utterance-level measures that speech-synthesis work reports are added to the final word's
harmonicity: a predicted naturalness score, UTMOS (the UTMOS22 strong learner, $1$ to $5$), and
speaker similarity, the cosine between WavLM-base-plus speaker-verification embeddings of a
steered recording and of the mean of its speaker's eight prompt recordings.

At the strengths we read, UTMOS stays within $0.2$ of the target condition's own recordings: $3.20$
against $3.30$ for CSM's steered rises, $3.35$ against $3.47$ for its falls, and $3.75$ against
$3.83$ and $3.85$ against $3.92$ in S2. At those strengths speaker similarity does not move ($0.898$
against $0.906$ for CSM's unsteered assertions; within $0.01$ in every cell). Past those strengths the two pushes
diverge. Toward a rise, harmonicity keeps increasing, because the pitch goes up, while UTMOS falls
steadily ($2.68$ at $4.5$ gaps in CSM). Toward a fall, flip rates rise from about $0.56$ at $1.5$
gaps to $0.80$ at $2.5$ in both systems, but harmonicity falls $4$ to $5$~dB below the target's at
the same time: the extra flips are paid for in voice quality, which is why we read the lower
strength. No listening test was run; these measures stand in for one.

\subsection{Voice quality conditioned on flipping}
\label{app:flipped}

Table~\ref{tab:flipped} splits each cell by the flip rule and reports the flipped subset against the
model's own unmodified recordings of the target condition, with intervals bootstrapped over
speakers. Runaway generations, which reach the frame cap without stopping, are excluded. Three of
the four cells overlap the target on every measure; the exception is S2's fall, whose harmonicity
interval does not reach the target's. In CSM's fall the \emph{unflipped} subset sits closer to the
target's harmonicity than the flipped one, which is how a cell mean can look like a match. Raising
the strength does not rescue the fall: at $2.5$ gaps CSM's flipped subset drops to voiced fraction
$0.741$ $[0.671, 0.801]$ and HNR $10.3$ $[8.9, 11.7]$~dB, and S2's to $0.734$ and $11.0$.

\begin{table}[!htbp]
\centering\small
\begin{tabular}{llrccc}
\toprule
system & subset & $n$ & voiced fraction & HNR (dB) & offset (ERB) \\
\midrule
\multicolumn{6}{l}{\emph{Push to a rise, $2.5$ gaps}} \\
CSM & flipped             & $49$ & $0.951$ $[0.926, 0.972]$ & $19.5$ $[17.5, 21.0]$ & $7.40$ $[6.66, 7.99]$ \\
CSM & target (unmodified) & $90$ & $0.927$ $[0.901, 0.951]$ & $17.7$ $[16.4, 19.1]$ & $6.58$ $[5.80, 7.35]$ \\
S2  & flipped             & $65$ & $0.887$ $[0.868, 0.903]$ & $19.4$ $[17.5, 21.1]$ & $7.25$ $[6.43, 8.09]$ \\
S2  & target (unmodified) & $90$ & $0.888$ $[0.877, 0.900]$ & $17.8$ $[16.2, 19.8]$ & $6.63$ $[5.87, 7.33]$ \\
\midrule
\multicolumn{6}{l}{\emph{Push to a fall, $1.5$ gaps}} \\
CSM & flipped             & $50$ & $0.883$ $[0.835, 0.922]$ & $13.9$ $[12.9, 14.7]$ & $4.19$ $[3.78, 4.73]$ \\
CSM & not flipped         & $39$ & $0.869$ $[0.829, 0.918]$ & $14.8$ $[13.3, 16.6]$ & $5.88$ $[5.27, 6.40]$ \\
CSM & target (unmodified) & $90$ & $0.857$ $[0.782, 0.915]$ & $15.3$ $[13.0, 17.5]$ & $4.69$ $[4.19, 5.27]$ \\
S2  & flipped             & $51$ & $0.816$ $[0.772, 0.859]$ & $13.9$ $[12.7, 14.9]$ & $4.10$ $[3.76, 4.59]$ \\
S2  & target (unmodified) & $90$ & $0.831$ $[0.779, 0.875]$ & $16.7$ $[14.4, 18.8]$ & $4.59$ $[4.04, 5.16]$ \\
\bottomrule
\end{tabular}
\caption{Voice quality of the steered recordings that flipped, against the target condition, with
$95\%$ intervals bootstrapped over speakers.}
\label{tab:flipped}
\end{table}

\FloatBarrier
\section{The corpus's tune shapes}
\label{app:shapes}\label{app:clusters}

\begin{figure}[!htbp]
\centering
\includegraphics[width=0.92\linewidth]{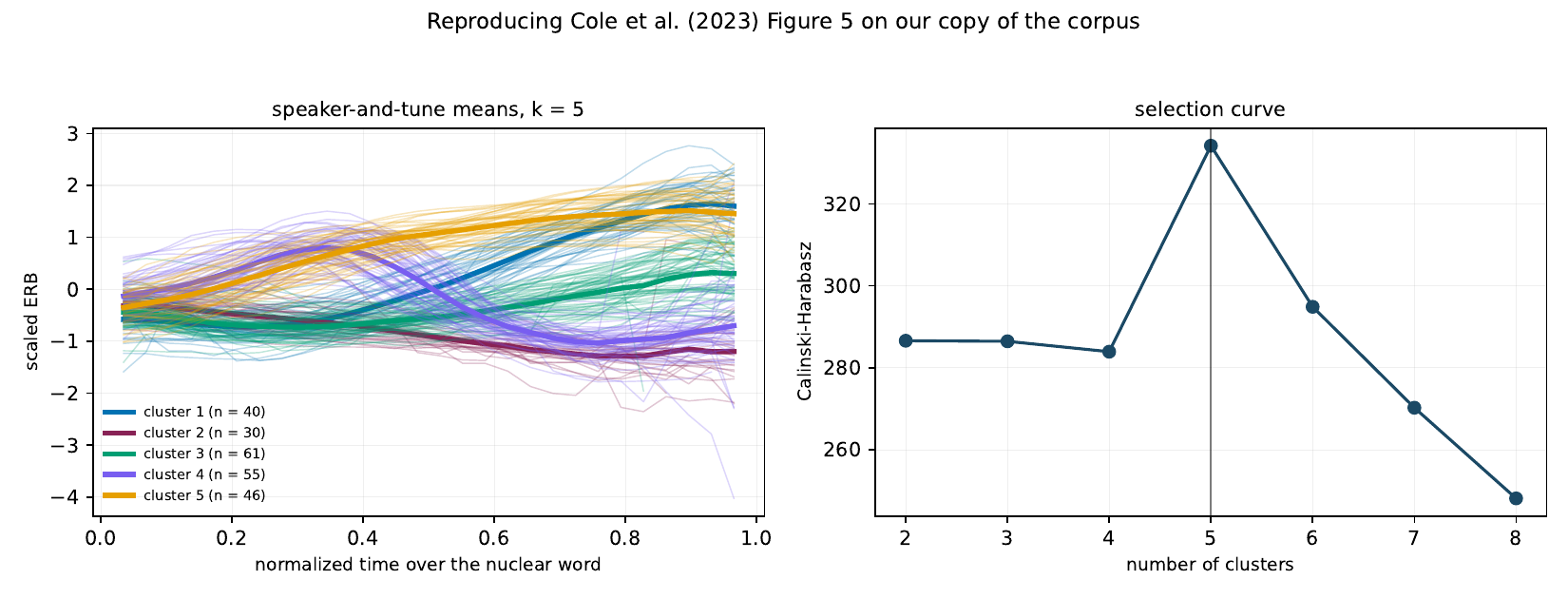}
\caption{The five nuclear-tune shapes of this corpus, reproducing Figure~5 of \citet{cole2023} on
the corpus recordings: one point per speaker and model tune, speaker-scaled ERB over
normalized time. \class{lhh} alone, \class{lll} alone, \class{lhl} with \class{llh}, \class{hlh}
with \class{hll}, and \class{hhl} with \class{hhh}. These centroids are what
Figure~\ref{fig:centroids} labels generated recordings against.}
\label{fig:cole}
\end{figure}

\citet{cole2023} cluster nuclear-tune contours from this corpus and find five groups. We repeat that
analysis on the corpus recordings, at their unit of analysis: one point per speaker and
model tune, $29$ speakers and eight model tunes giving $232$ points, in speaker-scaled ERB (each
speaker's contours centred on that speaker's mean and divided by their standard deviation), with
$k$-means (ten restarts, on the points two contours share) and the Calinski--Harabasz criterion.
One detail differs: they resample each contour over its own voiced extent, whereas we keep the
final word's own time base and leave the grid unvoiced outside the extent each recording phonates.
Calinski--Harabasz peaks at $k = 5$ and the partition agrees with the tune labels at adjusted Rand
$0.565$; against the labels grouped into the five classes below it agrees at $0.831$, with $213$
of the $232$ speaker-and-tune means ($92\%$) in their class's cluster. The five groups are the published ones (Figure~\ref{fig:cole}); running it their way
changes nothing material. The structure lives at the level of speaker-and-tune means: individual
trials of the same recordings select two groups, at adjusted Rand $0.157$.

\paragraph{Labelling a generated contour.} A generated recording is labelled by the nearest of
the five centroids, by the mean squared distance over the grid points both voice. Each generated
contour is first put in the centroids' space: centred on the mean, and divided by the spread, of
its own voice's unsteered recordings with both conditions pooled, the analogue of the speaker
scaling the centroids were built with. Steered recordings are scaled by the same unsteered
reference, so a change of level produced by the edit is kept. The label describes the contour's
shape in the corpus's terms, referring to the meaning-bearing prototypes derived from it.

\paragraph{Too little voicing.} A shape needs a contour. The pitch tracker finds too little
periodicity in some words --- devoicing and creak, which falling tunes invite and the voice
collapse of strong edits aggravates --- and grid points without $F_0$ are left empty. A generated word voiced
over less of its length than $99\%$ of the human productions ($52\%$ of the grid) is reported as having too little voicing and is not
labelled. This is a limit of $F_0$ extraction. The floor is taken from the human data
rather than set by hand, and not from their minimum, since three human recordings yield no $F_0$
at all. The exclusion is marginal where the paper reads its results: at most $3$ of $90$
recordings in any unsteered set or in the steered sets at the chosen strengths, in either system. It grows only for
strong pushes toward a fall, where the voice collapses (up to $69$ of $90$, for the \textsf{edge}
vector at $4.5$ gaps in CSM), and is drawn as its own bar in the strength-choice figures.

\begin{figure}[!htbp]
\centering
\includegraphics[width=\linewidth]{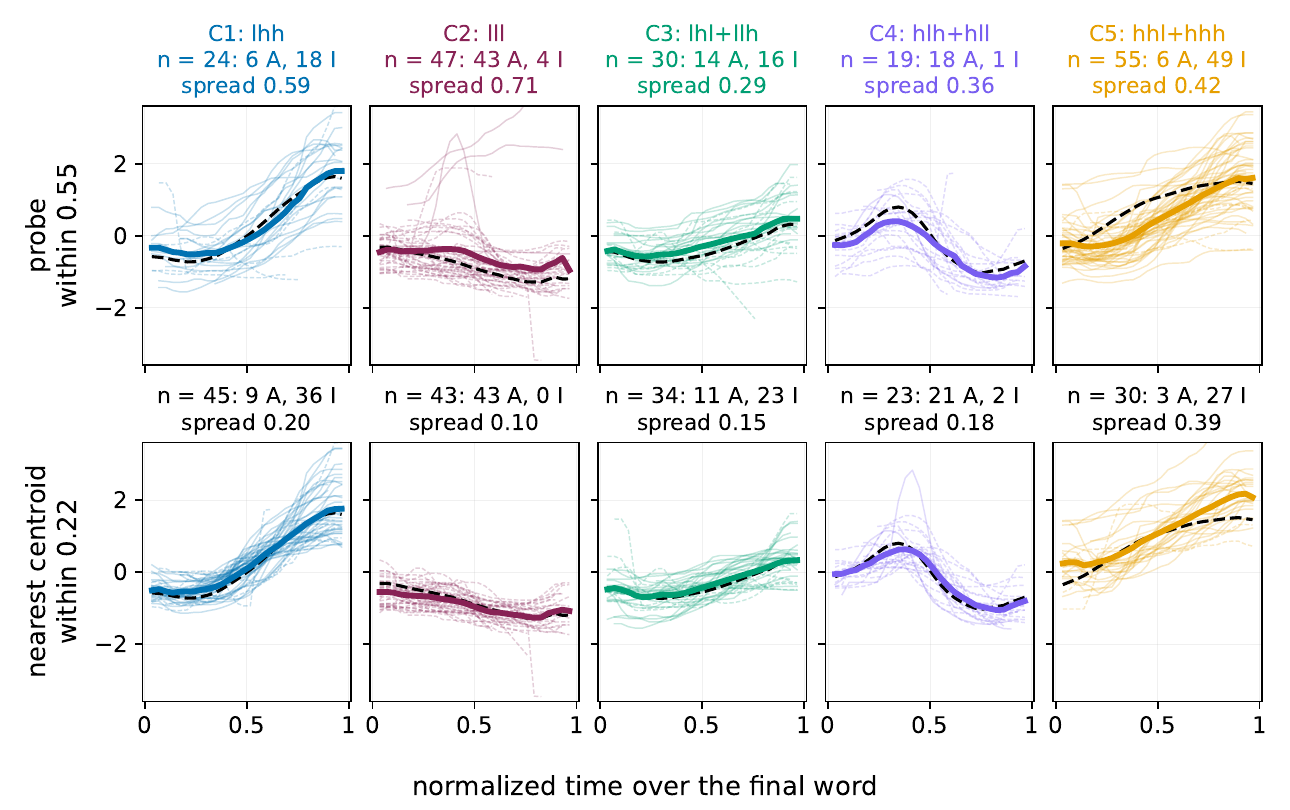}
\caption{CSM's unsteered recordings, assertions and questions pooled, as each labelling groups
them: one panel per class, recordings faint (dashed: assertions; solid: questions), the class mean
bold, and the human centroid dashed black; all in the centroids' scaled ERB. Each panel's title
gives its recordings by condition and its \emph{spread}, the mean squared distance to the class
mean; \emph{within} is the member-weighted spread as a share of the total. Words with too little
voicing are left out.}
\label{fig:labeller-csm}
\end{figure}

\begin{figure}[!htbp]
\centering
\includegraphics[width=\linewidth]{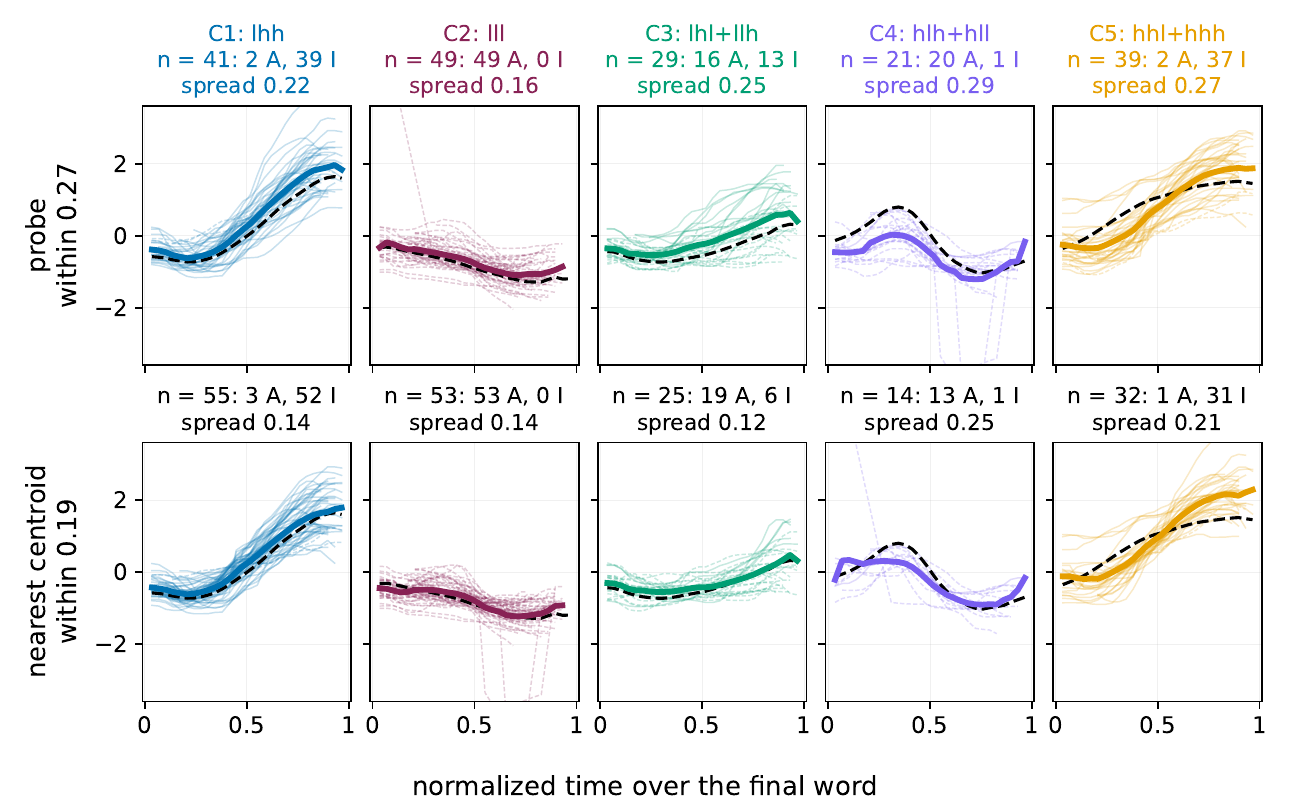}
\caption{S2's unsteered recordings as each labelling groups them, drawn as
Figure~\ref{fig:labeller-csm}.}
\label{fig:labeller-s2}
\end{figure}

\begin{figure}[!htbp]
\centering
\includegraphics[width=\linewidth]{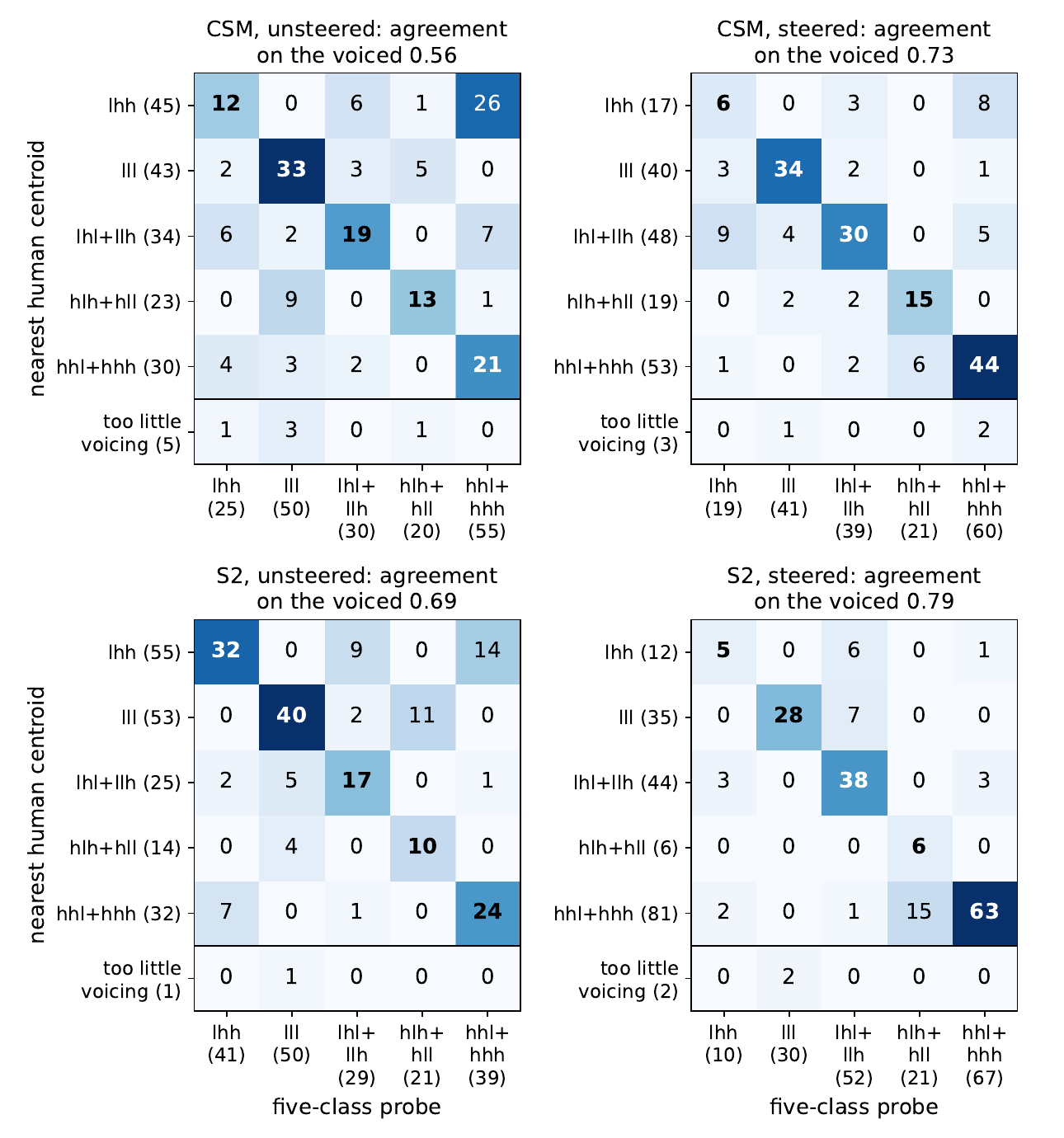}
\caption{The nearest human centroid (rows) against the five-class probe (columns), in counts, on the
unsteered recordings (left) and on those steered along the \textsf{tune} vector at the strengths
the paper reads (right: assertions to a rise at $2.5$ gaps, questions to a fall at $1.5$), with
class totals in brackets. The last row, below the rule, holds the words with too little voicing
for a contour, which the probe places in one of its five classes; agreement is the diagonal's share
of the voiced recordings above the rule.}
\label{fig:labeller-confusion}
\end{figure}

\paragraph{The alternative: a five-class probe.} The HuBERT probe of \S\ref{app:probes} labels the
same recordings by different machinery; its classes are the same five groups, so the two
labellings can be put in correspondence. Figures~\ref{fig:labeller-csm} and~\ref{fig:labeller-s2}
draw the unsteered recordings each labelling puts in each class, and
Figure~\ref{fig:labeller-confusion} tabulates the two against each other. On the voiced
recordings, the only ones both labellings can class, they agree on $0.56$ (CSM) and $0.69$ (S2)
of the unsteered recordings and on $0.73$ and $0.79$ of the steered ones at the strengths the
paper reads. Most disagreements are between \class{lhh} and the high plateau,
the two rising shapes, whose centroids differ mainly in when the rise begins. The two labellings
therefore give mostly compatible results, and the same account of where the steered recordings
land. Grouped by the nearest centroid, the recordings of a class are more alike: the within-class
spread is $0.22$ (CSM) and $0.19$ (S2) of the total, against $0.55$ and $0.27$ for the probe's
classes, $0.97$ for shuffled labels, and $0.21$ and $0.17$ for five-cluster $k$-means on the same
contours. Hence the classes the paper reports are groups of similar contour shapes.

The probe does not abstain on the words with too little voicing (the last row of
Figure~\ref{fig:labeller-confusion}): it always returns one of its five classes, and over every
punctuated in-loop recording it calls $85\%$ (CSM, $271$ of $319$ punctuated in-loop recordings) and $89\%$ (S2,
$188$ of $212$) of them \class{lll}. Where the voice collapses under a push toward a fall, the
probe would therefore report the models' most frequent assertion shape, and count a failure of
the voice as a successful fall.

\FloatBarrier
\section{Steering toward a tune without an orthographic cue}
\label{sec:hlh}\label{app:hlh}

\begin{figure}[!htbp]
\centering
\includegraphics[width=\linewidth]{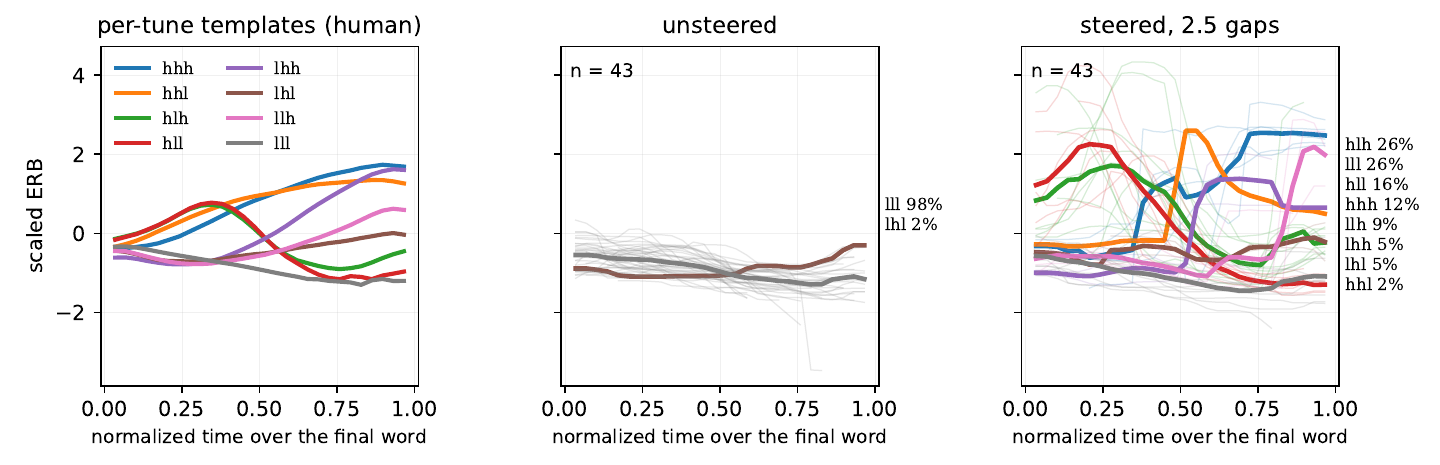}
\caption{The \class{hlh} pilot, CSM. \emph{Left:} the eight per-tune templates, in speaker-scaled
ERB over the final word. \emph{Middle:} the unsteered run of the $43$ assertions nearest
\class{lll}, each contour in the colour of its nearest template, bold lines the mean of each
colour; \emph{right:} the same recordings steered toward \class{hlh} at $2.5$ gaps. Percentages:
the share of the panel's voiced recordings nearest each template; words with too little voicing
are left out.}
\label{fig:hlh}
\end{figure}

The vector is a difference of means over the final word between the human recordings of
\class{hlh} and of \class{lll}, fitted on the backbone's pre-norm states of all ten speakers as in
\S\ref{sec:human} (gap $7.17$ state units; cosine with the punctuation vector $0.38$). It was
applied in the loop, as in \S\ref{sec:human}, to the CSM assertions whose unsteered run the
nearest-centroid labelling puts in \class{lll}: $43$ of the $90$, from all ten voices ($1$ to $6$
per voice). Because the five human clusters merge \class{hlh} with \class{hll}, each final-word
contour is labelled here by the nearest of eight per-tune templates: the mean over the corpus's
$29$ speakers of each speaker's mean contour of that tune, in speaker-scaled ERB, with generated
contours scaled by their voice's unsteered recordings (\S\ref{app:shapes}). Figure~\ref{fig:hlh}
draws the templates and the recordings at $0$ and $2.5$ gaps.

\begin{table}[!htbp]
\centering\small
\begin{tabular}{r r rrrrrrrr r}
\toprule
$\alpha$ (gaps) & $n$ & \class{hhh} & \class{hhl} & \class{hlh} & \class{hll} & \class{lhh} & \class{lhl} & \class{llh} & \class{lll} & excursion \\
\midrule
$0$   & $43$ & $0$  & $0$  & $0$  & $0$  & $0$ & $2$  & $0$ & $98$ & $-0.57$ \\
$1.5$ & $43$ & $5$  & $5$  & $2$  & $14$ & $5$ & $19$ & $2$ & $47$ & $-0.21$ \\
$2.5$ & $43$ & $12$ & $2$  & $26$ & $16$ & $5$ & $5$  & $9$ & $26$ & $-0.33$ \\
$3.5$ & $43$ & $21$ & $56$ & $2$  & $2$  & $5$ & $5$  & $5$ & $5$  & $0.59$ \\
$4.5$ & $40$ & $25$ & $68$ & $5$  & $2$  & $0$ & $0$  & $0$ & $0$  & $0.66$ \\
$6$   & $41$ & $15$ & $68$ & $0$  & $5$  & $0$ & $0$  & $0$ & $0$  & $0.29$ \\
$8$   & $36$ & $6$  & $75$ & $6$  & $0$  & $0$ & $0$  & $0$ & $0$  & $-0.06$ \\
\bottomrule
\end{tabular}
\caption{CSM assertions nearest \class{lll}, steered toward \class{hlh}: per cent of recordings
nearest each per-tune template, and mean excursion (ERB). $0$: the unsteered in-loop run. $n$
counts recordings that were generated and aligned; at $6$ and $8$ gaps, $12\%$ and $14\%$ of words
have too little voicing for a contour.}
\label{tab:hlh}
\end{table}

\section{Decomposing the tune: the edge and the accent}
\label{app:edge}\label{app:accent}

This appendix is for readers interested in the phonology. The tune vector of \S\ref{sec:human}
contrasts two tunes that differ in all three tone symbols. Two further vectors decompose it, fitted
the same way on groupings written with \class{x} for either tone:
\begin{center}\small
\begin{tabular}{llll}
\toprule
intervention & rising side & falling side & what it contrasts \\
\midrule
\textsf{tune}   & \class{lhh} & \class{hll} & the paradigmatic interrogative and assertive tunes \\
\textsf{edge}   & \class{xhh} & \class{xll} & high against low edge tone (phrase accent + boundary tone) \\
\textsf{accent} & \class{hxx} & \class{lxx} & the basic pitch-accent contrast (high against low), edge free \\
\bottomrule
\end{tabular}
\end{center}
The \textsf{tune} grouping uses $287$ recordings, \textsf{edge} $587$ and \textsf{accent} all
$1205$; the groupings are nested, so their comparison is not between independent samples.

\begin{figure}[!htbp]
\centering
\includegraphics[width=0.93\linewidth]{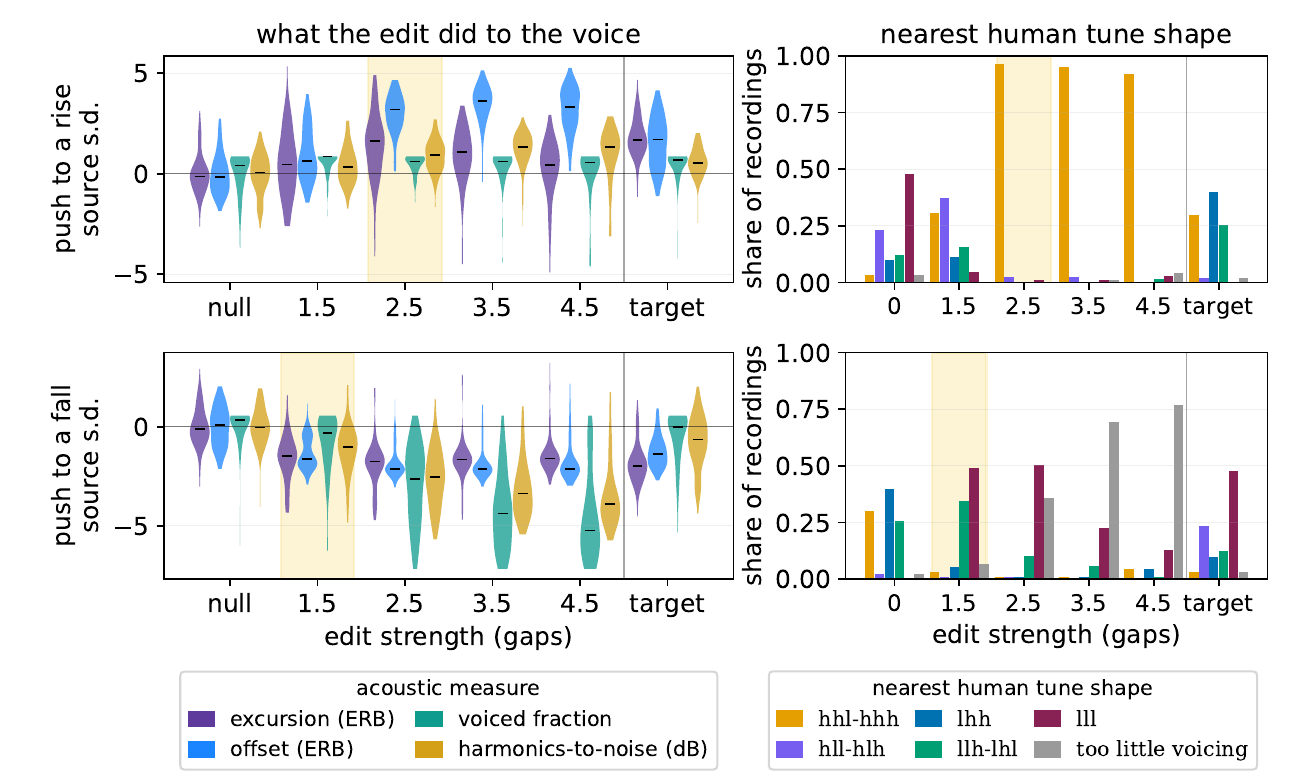}
\caption{Steering along the \textsf{edge} vector inside the loop, CSM, drawn as
Figure~\ref{fig:alpha}.}
\label{fig:alpha-edge}
\end{figure}

\begin{figure}[!htbp]
\centering
\includegraphics[width=\linewidth]{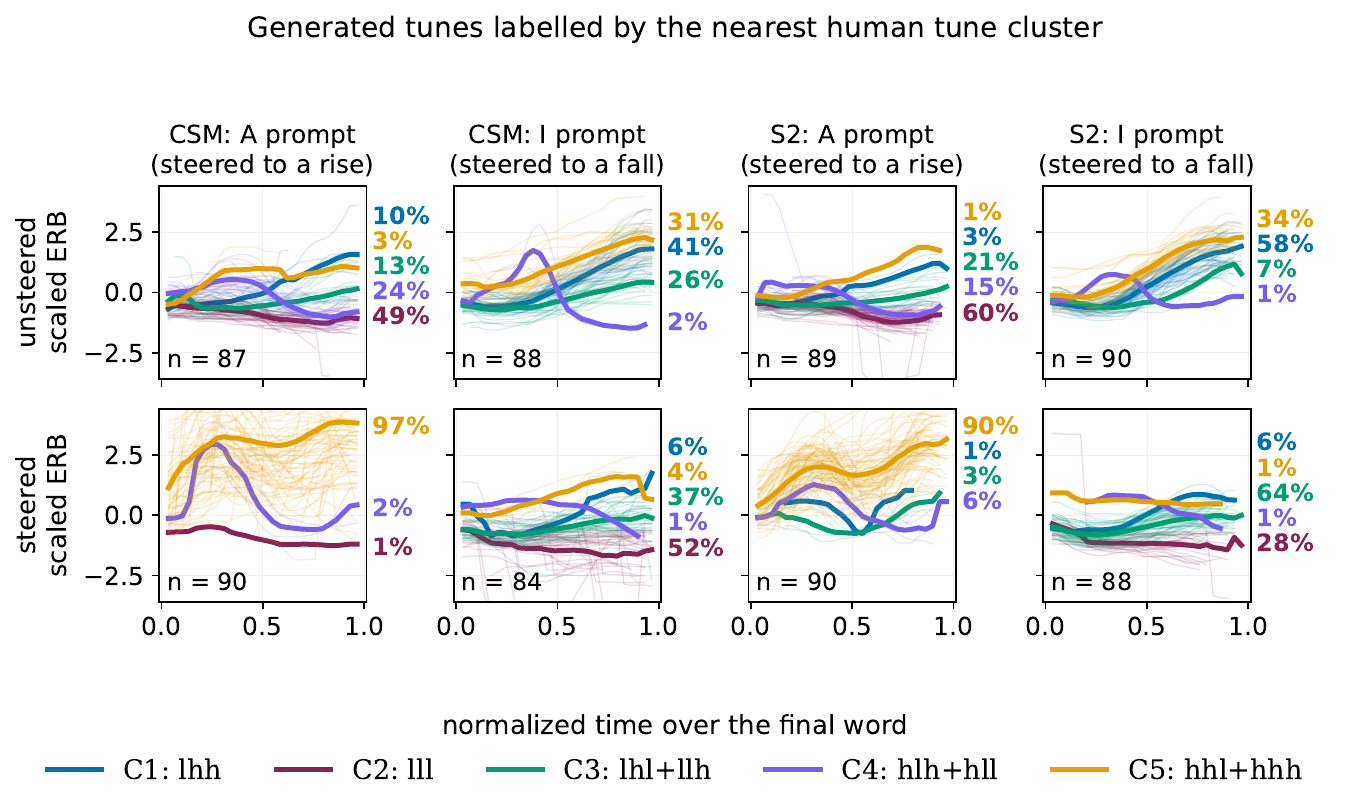}
\caption{Recordings steered along the \textsf{edge} vector labelled by the nearest human tune
cluster, drawn as Figure~\ref{fig:centroids}.}
\label{fig:centroids-edge}
\end{figure}

\begin{figure}[!htbp]
\centering
\includegraphics[width=0.93\linewidth]{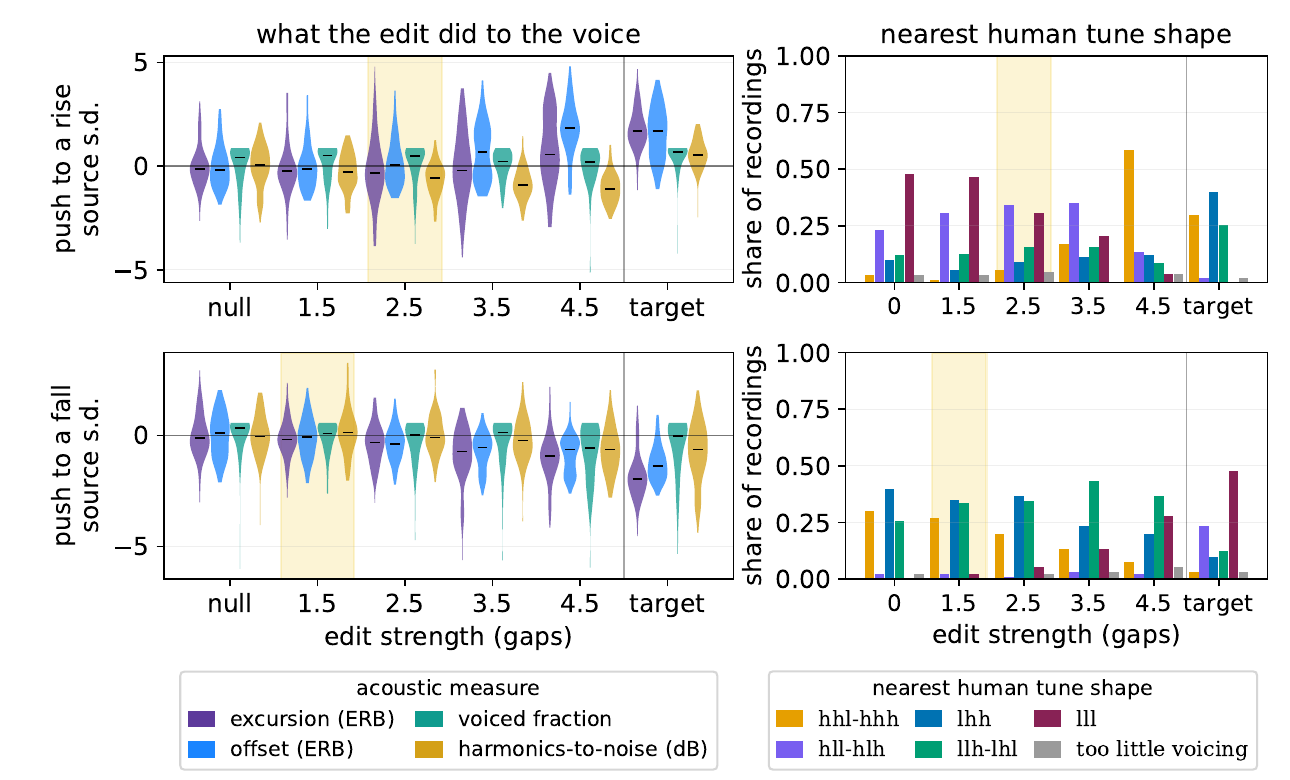}
\caption{Steering along the \textsf{accent} vector inside the loop, CSM, drawn as
Figure~\ref{fig:alpha}.}
\label{fig:alpha-accent}
\end{figure}

\begin{figure}[!htbp]
\centering
\includegraphics[width=\linewidth]{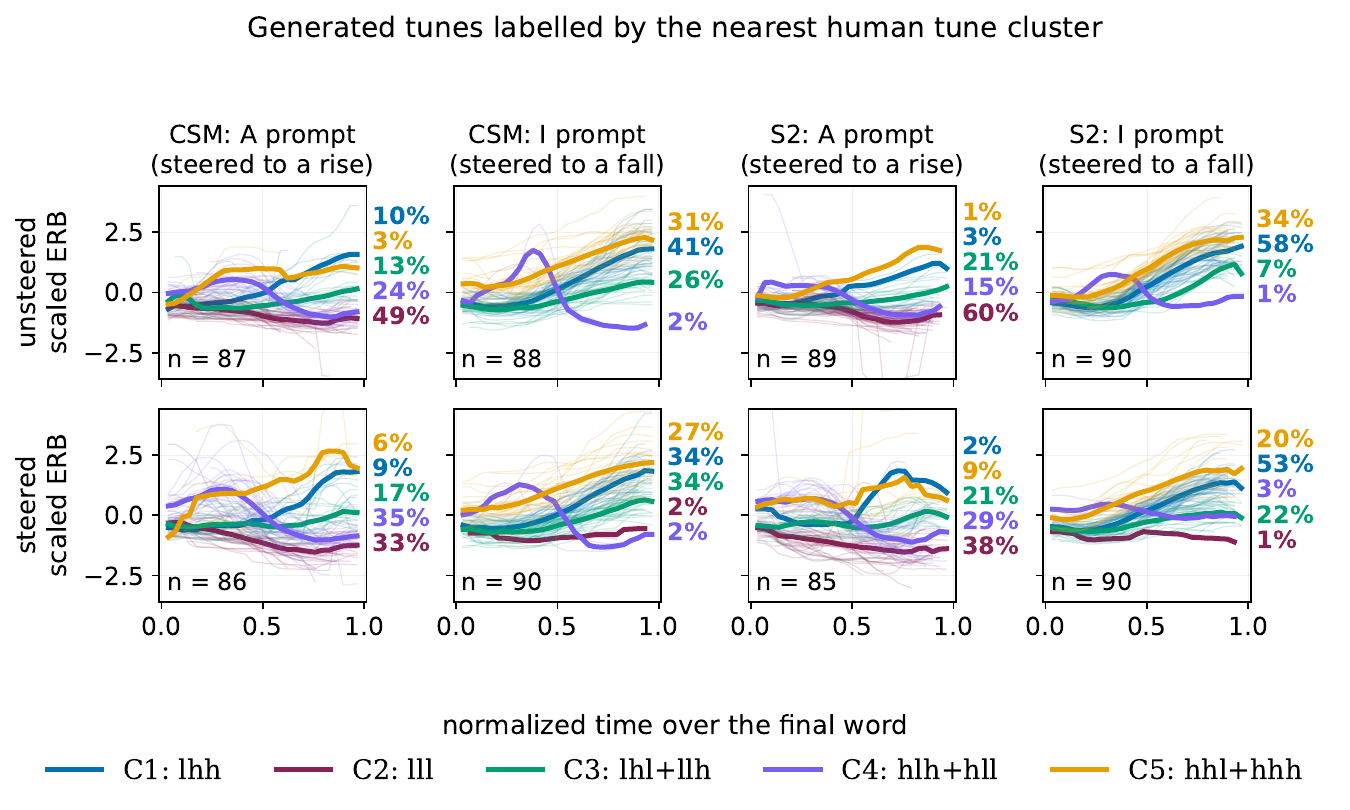}
\caption{Recordings steered along the \textsf{accent} vector labelled by the nearest human tune
cluster, drawn as Figure~\ref{fig:centroids}.}
\label{fig:centroids-accent}
\end{figure}

\paragraph{The contrast is an edge-tone contrast.} The \textsf{edge} vector reproduces the
\textsf{tune} vector's effect throughout (Table~\ref{tab:flips}, Figures~\ref{fig:alpha-edge}
and~\ref{fig:centroids-edge}), has the higher cosine with the punctuation direction ($0.78$ against
$0.66$ in CSM, $0.80$ against $0.44$ in S2) and separates the model's own two conditions better
($0.74$ against $0.71$; $0.79$ against $0.63$). The \textsf{accent} vector does none of this:
cosine $0.28$, a separation of the model's own two conditions along it of $2.85$ state units
against $7.10$ and $8.34$ for the other two, and flip rates that do
not clear the no-edit baseline until $3.5$ gaps, its effect at lower strengths being indistinguishable
from resampling (Figures~\ref{fig:alpha-accent} and~\ref{fig:centroids-accent}). The
assertion--question contrast these models encode is carried by the phrase accent and boundary
tone, with the pitch accent contributing little, which agrees with where the contrast sits
acoustically (Figure~\ref{fig:final-word}) and with the phonology of the two tunes.

S2 departs from this in one respect. Its \textsf{accent} vector is more aligned with the
punctuation direction than CSM's ($0.58$ against $0.28$), separates S2's own two conditions better
(mass-mean accuracy $0.73$ against $0.62$), and at $3.5$ and $4.5$ gaps flips $0.53$ and $0.74$ of
S2's questions toward an assertion, against $0.33$ and $0.34$ in CSM (Table~\ref{tab:flips}).
Since neither system's unsteered conditions differ at the final word's onset, where the accent
sits (Figure~\ref{fig:final-word}), a vector fitted on the pitch accent alone can move the flip
read-out only through the edge. If it affected the pitch accent only, it should leave the edge
alone; that it does not suggests that S2 represents pitch accent and edge tone jointly, in a way
that conflicts with the linguistic distinction between them and with human productions, in which
they vary independently.

This conclusion has a boundary. The flip read-outs are edge instruments: the excursion is the last
fifth of the word minus the first, the velocity is that quantity differentiated, and the edge probe
is an \class{xll}-versus-\class{xhh} classifier. An intervention that moved the pitch accent and
left the edge alone would score near zero on all three. The accent probe, which could break the
circle, is uninformative on generated material (\S\ref{app:probes}). The finding is that the
contrast \emph{as measured by movement at the edge of the final word} is carried by the edge tones;
it is not evidence that the accent is absent from the state, and the \textsf{accent} vector still
separates held-out human recordings at $0.69$ (CSM) and $0.71$ (S2), so it is a real direction
carrying real information.

\paragraph{The neutral prompt.} A third prompt condition, in which the sentence ends in no
punctuation at all, was run as an ablation. Without a final mark CSM fails to emit its
end-of-utterance frame on $7.8\%$ of generations against $1.1\%$ with one; S2, whose prompt template
supplies an explicit stop token, is unaffected. Those recordings carry speech their transcript does
not, so the aligner refuses them, and they are excluded. Within what remains, the tune moves as it
does under the punctuated prompts.
\FloatBarrier
\end{document}